\documentclass[twocolumn]{aa}

\usepackage{graphicx}
\usepackage{txfonts}
\begin{document}

   \title{Phase resolved spectrum of the Crab pulsar from {\it{NICER}}}

   \author{M. Vivekanand \inst{}}

   \institute{No. 24, NTI Layout 1\textsuperscript{st} Stage, 
              3\textsuperscript{rd} Main, 1\textsuperscript{st} Cross, 
	      Nagasettyhalli, Bangalore 560094, India. \\
   \email{viv.maddali@gmail.com}}

   \date{}

 
  \abstract
   {
   The high energy emission regions of rotation powered pulsars are studied using folded light curves 
   (FLCs) and phase resolved spectra (PRS).
   }
   { This work uses the {\it{NICER}} observatory to obtain the highest resolution FLC and PRS of the Crab 
   pulsar at soft X-ray energies.
   }
   {
   {\it{NICER}} has accumulated about $347$ ksec of data on the Crab pulsar. The data are processed 
   using the standard analysis pipeline. Stringent filtering is done for spectral analysis. The 
   individual detectors are calibrated in terms of long time light curve (LTLC), raw spectrum and 
   deadtime. The arrival times of the photons are referred to the solar system's barycenter and the 
   rotation frequency $\nu$ and its time derivative $\dot \nu$ are used to derive the rotation phase 
   of each photon.
   }
   {
   The LTLCs, raw spectra and deadtimes of the individual detectors are statistically similar; the 
   latter two show no evolution with epoch; detector deadtime is independent of photon energy. The 
   deadtime for the Crab pulsar, taking into account the two types of deadtime, is only $\approx 7$\% 
   to $8$\% larger than that obtained using the cleaned events. Detector $00$ behaves slightly 
   differently from the rest, but can be used for spectral work. The PRS of the two peaks of the 
   Crab pulsar are obtained at a resolution of better than $1/512$ in rotation phase. The FLC very 
   close to the first peak rises slowly and falls faster. The spectral index of the PRS is almost 
   constant very close to the first peak.
   }
   {
   The high resolution FLC and PRS of the {{peaks}} of the Crab pulsar provide important 
   constraints for the formation of caustics in the emission zone.
   }


   \maketitle
%

\section{Introduction}

Although rotation powered pulsars (RPPs) were first discovered at radio wavelengths, they
emit mainly at the much higher X-ray and $\gamma$-ray energies. So observations at these 
high energies are required to understand their emission mechanism. Further, the radio
emission of RPPs is believed to be due to a coherent radiation mechanism, which is relatively 
difficult to study. In contrast, the high energy emission is relatively simple -- ultra 
relativistic charges traveling along magnetic field lines and emitting very narrow beams 
of radiation along the direction of their momentum. Therefore a realistic magnetic field 
line geometry (e.g. dipole field with swept back field lines close to the light cylinder) 
is sufficient to qualitatively reproduce the observed folded light curve (FLC) of RPPs. 
Such work gained speed after the launch of the {\it{Fermi}} Gamma-Ray Space Telescope in 2008 
\citep{Atwood2007}; see also \cite{Harding2016}.

The earliest magnetic field geometry used for modeling was {{the static dipole}}. More realistic 
magnetic fields were explored as time progressed -- {{the vacuum retarded dipole}}, the aligned 
force-free magnetosphere, in which the magnetic and rotation axes are parallel, and the 
oblique force-free magnetosphere, in which they are not parallel. The force-free magnetospheres
do not have any particle acceleration by definition. Therefore dissipative magnetospheres
were invoked, that had a finite conductivity and allowed for some acceleration. However,
the dissipative models did not model self-consistently the source of particles that provide
the conductivity. This led to modeling of pulsar magnetospheres using particle-in-cell 
codes, which are still in their infancy; see Harding 2016 and references therein.

The magnetospheric models were coupled with different sites of particle acceleration,
known as ``gaps'', in which space charge deficiency leads to residual electric fields 
and therefore to particle acceleration. The popular gaps are the polar cap gap (PC), the 
outer gap (OG), the slot gap (SG) and its associated two pole caustic (TPC) model, and 
the separatrix layer (SL) model (see Harding 2016 and references therein).

Most of the above work was done at $\gamma$-ray energies. While each model explained 
qualitatively the FLCs of some RPPs, none of them explained the FLCs of all {\it{Fermi}} RPPs 
\citep{Bai2010, Romani2010, Kalapotharakos2014, Harding2016, Pierbattista2016}. Further, 
quantitative comparison of the observed and modeled FLCs (in terms of $\chi^2$) was 
often quite poor \citep{Harding2016}. However, several of the above models were able to 
reproduce the typical double peaked FLC even from a single magnetic pole of the RPP. 
They were able to reproduce the sharp peaks in terms of caustics, in which photons from 
different parts of the emission region arrived at the same rotation phase simultaneously,
due to a combination of light travel time and special relativistic aberration 
\citep{Cheng1986}. However, analysis of the phase resolved spectrum (PRS) has lagged 
behind the analysis of the FLC in most of these models, until recently 
\citep{Cheng2000, Brambilla2015}.

The most popular RPP for such work is the Crab pulsar. It is a very luminous source, so 
it has been observed right from radio to $\gamma$-ray energies. Further, it is probably 
the only RPP for which the FLCs at most energies are not only similar in 
appearance, but are also aligned in rotation phase (except for the radio precursor; 
Abdo et al. 2010). This implies that the geometry of the emission region is common for 
most energies in the Crab pulsar. Therefore modeling its FLC at X-ray energies 
may be as useful as modeling it at $\gamma$-ray energies. This work obtains the high 
phase resolution (large number of bins per rotation period) FLC and PRS of the Crab 
pulsar at soft X-ray energies {{($1 - 10$ keV)}} using the Neutron star Interior 
Composition Explorer ({\it{NICER}}) satellite observatory \citep{Arzoumanian2014, 
Gendreau2016},  {{ which is the main result of this work. Here the focus will be 
mainly on the formation of caustics or cusps in the FLC of the Crab pulsar at soft
X-ray energies.}}

{{
\cite{Cheng1986} and \cite{Romani1995} were one of the earliest theorists to demonstrate the 
formation of sharp peaks in the $\gamma$ ray FLC of RPPs. Using a static dipole magnetic field, 
\cite{Cheng1986} showed that sharp peaks can form in the FLC due to photons from a large portion 
of the outer gap (OG) arriving at a common rotation phase (see their Figure $11$). These so called 
``caustics'' or ``cusps'' form when one takes into account the combined effect of the special 
relativistic aberration and light travel time on a photon. \cite{Romani1995} demonstrated the same using 
the more sophisticated vacuum retarded dipole (VRD) magnetic field. Clearly the sweep back of the 
magnetic field lines in a VRD did not affect the process of caustic formation. These developments 
enabled the explanation of the well separated double peaks in $\gamma$ ray FLCs in terms of 
emission from a single magnetic pole of the RPP.

Then the {\it{Fermi}} Gamma-Ray Space Telescope obtained high quality $\gamma$ ray FLCs of several RPPs. 
This enabled statistical testing of more refined theoretical models of the formation of FLC.
\cite{Romani2010} tested three types of magnetospheric structures and two locations of the emission 
zone. The three magnetospheric structures were static magnetic field, modified VRD which they label 
``atlas'' field, and ``pseudo force free'' (PFF) magnetosphere, which contains charges, but no 
currents. The two locations were the two pole caustic (TPC) and the outer gap (OG). They concluded
that statistically OG was preferable to TPC. However, their best-fit values of $\alpha$, the angle 
between rotation and magnetic axis, and $\zeta$, the angle between rotation axis and line of sight, 
differ significantly from independently known values for some RPPs. In the work of \cite{Romani2010} 
caustics are caused by piling up in phase of emission from different regions (altitudes) of the 
magnetosphere.

\cite{Bai2010} use only the three dimensional force free magnetosphere (FF), which by definition 
has no particle acceleration; and they explore the TPC and OG locations. The FF magnetosphere has 
a conducting plasma, and the charges and currents are solved self-consistently. They conclude that 
both the TPC and OG models are unsuitable, but favor a modified version of OG which they call the 
``separatrix layer'' gap (SL). In this model caustics form by an entirely different process -- 
that of radiation from the same magnetic field line piling up in phase. This is possible because 
outside the light cylinder the magnetic field lines are almost radial, in the so called ``split 
monopolar'' geometry. They have named this kind of caustic formation the ``sky map stagnation'' 
(SMS) effect. These caustics form in the outer magnetosphere.  Each magnetic pole gives rise to 
one peak in the FLC and the rise of flux to the peak as a function of rotation phase is usually 
faster than its drop after the peak.

There are two important aspects of caustic formation. First, it depends sensitively upon the 
geometry of the magnetic field lines, since that determines the amount of aberration and light 
travel time \citep{Bai2010}. Second, as emphasized by \cite{Bai2010}, the magnetic field structure at the light 
cylinder determines the shape of the polar cap on the surface of the neutron star, which in turn
determines the shape and extent of the region of emission of photons, which affects the formation
of caustics.

\cite{Kalapotharakos2014} use a dissipative magnetosphere, using different values of conductivity
$\sigma$; when $\sigma = 0$, the situation is similar to the VRD magnetosphere, while $\sigma = 
\infty$ implies the FF magnetosphere. For small values of $\sigma$ they find that the emission is 
mainly from the inner magnetosphere and the peaks in the FLC are not sharp. For large values of 
$\sigma$ the emission is mainly from the outer magnetosphere and the peaks can be sharp. For
very large values of $\sigma$ the emission is entirely from the outer magnetosphere and is very
non-uniform, so complex FLCs are obtained. They favor a hybrid model that is FF inside the light 
cylinder and dissipative outside (FIDO). In terms of caustic formation, the FIDO model is 
essentially two pole, like in the SG model, but without emission at low altitudes or at off peak 
phases, like in the OG model.

The above is a brief summary of the mechanism of caustic formation in high energy FLCs of RPPs.
The results of this work can address some questions regarding caustic formation.
}}
To achieve this, two aspects have to be kept in mind. First, {\it{NICER}} consists of $56$ 
co-aligned detectors, each of which is essentially a small X-ray telescope. This clever
design avoids (for most sources) two of the difficulties commonly faced by X-ray 
observatories -- pile-up of photons and detector deadtime. Correspondingly, this design 
requires the calibration of $56$ detectors in terms of their long time light curves 
(LTLCs) and their spectral properties. This has been done in this work for the $49$ 
useful detectors of {\it{NICER}}.

Second, to obtain FLC and PRS at high phase resolution, one needs highly accurate 
rotation frequency $\nu$ and its time derivative $\dot \nu$ as a function of epoch, 
since these parameters of the Crab pulsar change significantly with time and since the 
{\it{NICER}} observations of the Crab pulsar occurred over a duration of about $2.35$ years. 
The $\nu$ and $\dot \nu$ have been obtained self consistently from the {\it{NICER}} data  
itself and only that data has been used for which the error in the derived rotational 
phase of each photon is less than $1/2048$ of a period. This enables one to obtain the 
FLC and PRS of the Crab pulsar at a resolution of $1024$ bins per period, and certainly
better than $512$ bins per period. {{ The details of how these $\nu$ and $\dot \nu$ 
are obtained, the treatment of the Crab pulsar glitch of $2017$ November, and the
consistency with the radio ephemeris are explained in \cite{Vivekanand2020}. Typically 
the accuracy of $\nu$ obtained by NICER is $0.0009$ $\mu$Hz, with a standard deviation 
of $0.0006$ $\mu$Hz. The typical accuracy of the $\dot \nu$ is $4 \times 10^{-14}$ Hz/s, 
with a standard deviation of $3 \times 10^{-14}$ Hz/s.}}

Section $2$ discusses the observations and analysis of this work. The next three sections 
describe the calibration of the $49$ useful detectors of {\it{NICER}}. Sections $6$ and $7$ 
presents the spectrum for the Crab pulsar. The last section discusses the implications 
of this work.

\section{Observations and analysis}

{\it{NICER}} consists of $56$ co aligned detectors \citep{Arzoumanian2014, Gendreau2016, 
Prigozhin2016}. Its operating range is $0.2 - 12$ keV with a peak collecting area of $1900$ 
cm$^{2}$ at $1.5$ keV and below. Time resolution is $0.1$ microsecond ($\mu$s). At the 
time of doing this work, {\it{NICER}} had observed the Crab pulsar on $72$ different days, 
starting from $2017$ Aug $5$ to $2020$ Apr $27$, which is a duration of $2.73$ years, with 
multiple observations on $8$ of those days, resulting in $80$ observation identity numbers 
(ObsID). Ten of these ObsIDs had live times of less than $100$ sec, so they were not 
analyzed. Four of them were ignored for reasons specified in \cite{Vivekanand2020}, which 
gives details of the earlier observations spanning $1.72$ years and their analysis. The 
{{ analysis of this work }} is slightly different because of the enhanced sensitivity 
requirement for spectral work.

The data were analyzed using {\it{NICER}} version $2020$-$04$-$23$\_V$007$a software included in 
the HEAsoft distribution 6.27.2; the calibration setup was CALDB version XTI($20200722$). 
The analysis for each ObsID begins with the pipeline tool {\it{nicerl2}} with the following 
parameters: angle between pointing and earth limb ELV $= 30^\circ$, and angle between pointing 
and bright earth BR\_EARTH $= 40^\circ$ \citep{Stevens2018, Bogdanov2019, Guillot2019, Miller2019, 
Trakhtenbrot2019}; magnetic cut off rigidity COR\_RANGE $\ge 4.0$ \citep{Bogdanov2019, 
Guillot2019, Malacaria2019, Riley2019, Trakhtenbrot2019, vandenEijnden2020}; and undershoot 
count rate UNDERONLY\_RANGE $= 0$-$50$; the rest of the parameters have default values 
\citep{Stevens2018, Bogdanov2019}. Next, the tool {\it{nimaketime}} is used to create good 
time intervals (GTI), using additionally the following two criterion: angle between 
pointing and sun SUN\_ANGLE $> 90^\circ$ \citep{Bogdanov2019}; and observations must be done when 
{\it{NICER}} in not in sun light, SUNSHINE $= 0$ \citep{Hare2020}. 

Next, the tool {\it{fselect}} is used to exclude data of detectors $14$, $34$ and $54$, 
whose data is known to be problematic \citep{Bogdanov2019, Ray2019}. Coupled with the detectors
$11$, $20$, $22$ and $60$ that are permanently switched off at {\it{NICER}}, this leaves $49$ 
out of the original $56$ detectors for our analysis. Then, the original GTIs of the 
event file are replaced by the GTIs obtained above (if required, the FITS file extension 
name EXTNAME has to be changed back to GTI\_FILT). Then {\it{fselect}} is used to filter 
out photon events outside the GTIs, using the {\it{gtifilter}} option. Next the LTLC 
is obtained using the tool {\it{extractor}} with a bin size of $2.5$ sec. This is 
used to exclude durations of observations in which the mean count rate is unusually low 
or high; this is done by editing the GTI extension of the event file and filtering using 
{\it{fselect}} with the {\it{gtifilter}} option.

Then, {\it{fselect}} is used to extract the data of each individual detector and 
{\it{extractor}} is used to obtain its LTLC. The tool {\it{fstatistic}} is used on 
the LTLC to obtain basic statistics of the count rate of each detector -- the mean 
value, its standard deviation, the minimum and maximum values. Unusually low or high
values indicate flaring, satellite not exactly on source, etc.; these durations are 
also excluded.

High background counts are identified by extracting a LTLC of bin size $8$ sec in
the energy range $12 - 15$ keV and excluding those durations in which the count rate
is statistically higher than $1.0$ \citep{Bult2018}. Details are given in Appendix A.
{{ {\it{NICER}}'s background is believed to be typically less than $1$ count/s/keV. 
It is obtained by processing data from several {\it{RXTE}} background regions. 
Background region $5$ of {\it{RXTE}} has an approximately power-law spectrum that is 
$< 1.0$ counts/s/keV at $0.4$ keV and $< 0.1$ counts/s/keV above $1.4$ keV 
\citep{Keek2018}. In the top panel of their Figure $2$, the background decreases to 
$\approx 0.01$ counts/s/keV at $10$ keV, while the flux of the Crab nebula plus pulsar
is $3.7$ counts/s/keV at that energy, which can be derived from Figure~\ref{fig4}. 
Thus the background is $\approx 370$ times weaker than the Crab {{nebula plus}} pulsar 
at $\approx 10$ keV. }}

High particle background due to geomagnetic storms are accounted for by the space weather
parameter Kp \citep{Ludlam2019, Miller2019}, which is required to be below the value $5$.
This is done by first converting the START and STOP times in the event files from the TT
timescale to UTC dates and times and then checking the Kp values during the observation 
in the corresponding yearly WDC files\footnote{ftp://ftp.gfz-potsdam.de/pub/home/obs/kp-ap/wdc/yearly/}.
Details are given in Appendix B.

Finally, the event epochs that remain after the above stringent filtering are referenced 
to the solar system barycenter using the {\it{barycorr}} tool, using the JPL ephemeris 
DE430, with the position of the Crab pulsar at that epoch as input. The $\nu$ and $\dot 
\nu$, and their reference epoch for each ObsID are then used to compute the rotation 
phase of each photon, which is written into the event file. \cite{Vivekanand2020} 
describes most of the above analysis in a different context. Later sections of this work 
describe the specific analysis required for the corresponding sections.

The filtering above reduced the total live time by $\approx 60$\%, to $\approx 139$ ksec,
leaving only $29$ ObsIDs with sufficient data to analyze, that also had their 
$\nu$ and $\dot \nu$ estimated by \cite{Vivekanand2020}, who does not implement this 
stringent filtering, who thus had many more photons for obtaining $\nu$ and $\dot \nu$. 
For those groups of ObsIDs that had sufficient number of photons in spite of the stringent 
filtering, it was possible to estimate $\nu$ and $\dot \nu$, which were consistent with 
those derived by \cite{Vivekanand2020}.

\section{Long time light curve (LTLC)}

\begin{figure}[h]
\begin{center}
\includegraphics[keepaspectratio=true,scale=1.0,width=9.5cm]{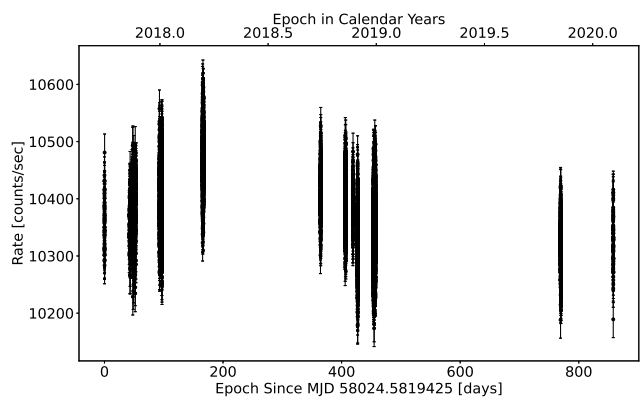}
\end{center}
\caption{
	LTLC of $49$ useful detectors of {\it{NICER}} combined, over $\approx 2.35$ years, with a bin 
	size of $10$ seconds.
	}
\label{fig1}
\end{figure}

Figure~\ref{fig1} shows the LTLC of the Crab {{nebula plus}} pulsar using the $49$ useful detectors of {\it{NICER}}, 
in the energy range $0.2 - 12$ keV, for the $2.35$ years of observations used in this 
work, with a bin size of $10$ sec. This was obtained using the tool {\it{extractor}}. The mean 
count rate is {{
$10384$ per second with a standard deviation of $75$, which is $0.72$\% of the
mean. The maximum and minimum values in Figure~\ref{fig1} are $10610.0$ and $10173.5$ counts
per second, which implies that the peak-to-peak variation of the LTLC is about $\approx 
(10610.0 - 10173.5) / 10384 \approx 4.2$\%. This is consistent with the recent measurement of 
the Crab {{nebula plus}} pulsar's LTLC by the {\it{NuSTAR}} and {\it{Swift/BAT}} 
missions\footnote{(https://iachec.org/wp-content/presentations/2019/sessionI\_-Madsen.pdf)}.
This is also consistent with the findings in the recent past, that the reported decrease
in the Crab nebula's X-ray flux by $\approx 7$\% over $\approx 2$ years starting from MJD 
$54690$ (mid $2008$), as reported by \cite{Wilson-Hodge2011}, is no longer observed.\footnote{https://ntrs.nasa.gov/api/citations/20120015012/downloads/-20120015012.pdf}\footnote{http://www.iucaa.in/iachec/talkmaterials//66/IACHEC-02-29-2016\%20-\%20Case.pptx}
In fact \cite{Wilson-Hodge2011} themselves stated that they cannot predict if the flux decline will 
continue in the future. This result is also confirmed in Fig. $1$ of \cite{Kouzu2013}.
}}

\begin{figure}[h]
\begin{center}
\includegraphics[keepaspectratio=true,scale=1.0,width=9.5cm]{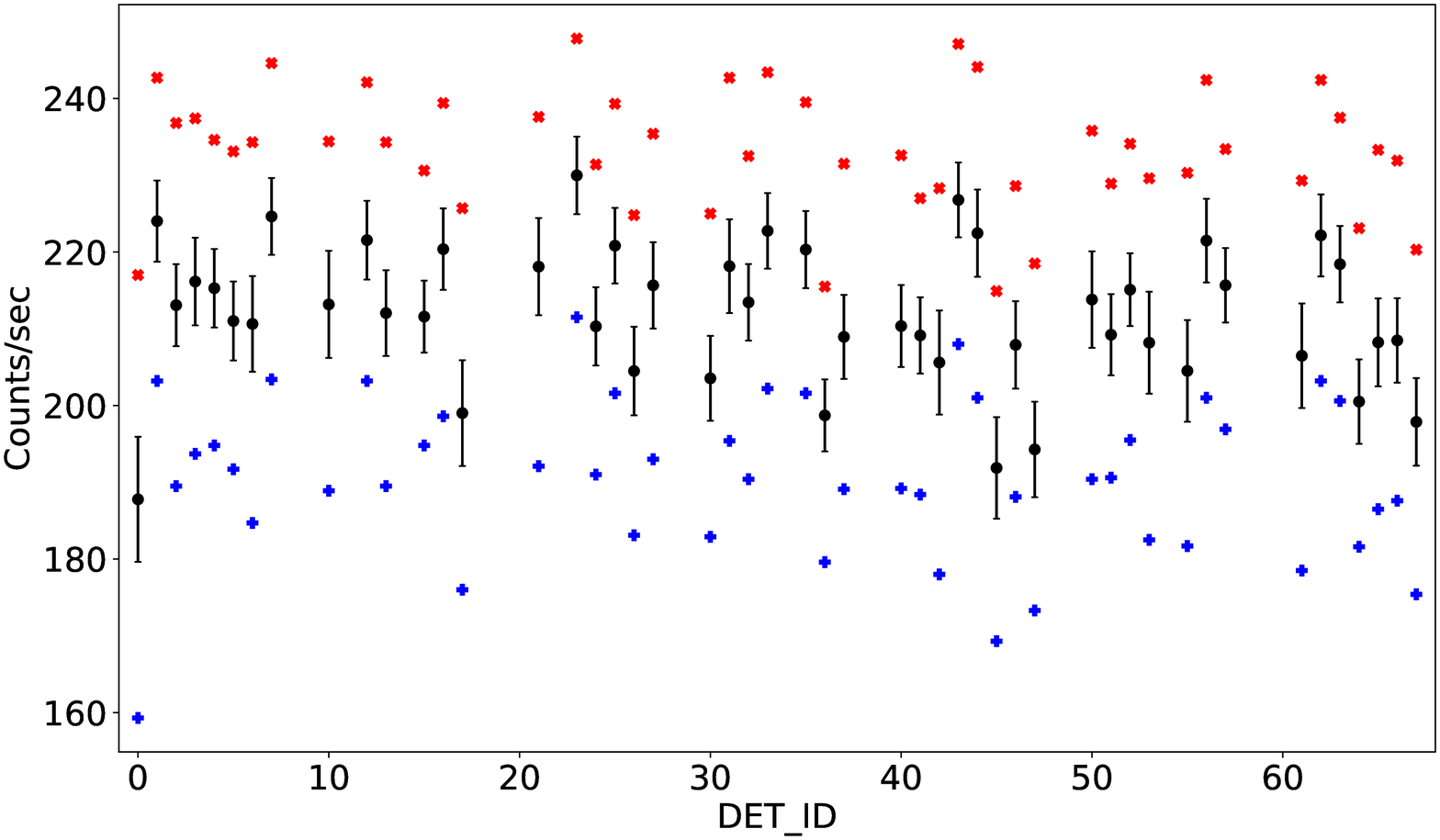}
\end{center}
\caption{
	Statistics of the LTLCs of the $49$ individual detectors of {\it{NICER}}. {{ The black dots,
	red crosses and blue pluses are the mean count rate, the maximum and the minimum values, 
	respectively. The error bars represent the standard deviation and not the error on the mean.}}
	}
\label{fig2}
\end{figure}

Figure~\ref{fig2} shows the statistics of the $49$ individual LTLCs. The dots with error bars 
show the mean count rate for each detector over the $2.35$ year duration. The crosses and the 
pluses represent the maximum and minimum values, respectively. All except detector $00$ show 
statistically similar values. Detector $00$ has a significantly lower mean count rate and a 
standard deviation slightly higher than the others. {{ For example, the mean count rates of 
detectors $00$ and $01$ are $187.78$ and $224.02$, respectively, while their standard 
deviations are $8.15$ and $5.29$.}}
The minimum values in Figure~\ref{fig2} are 
typically $-3.8(3)$ standard deviations below the mean, the error in the last digit given in 
the parenthesis; the maximum values are typically $+3.8(3)$ standard deviations above the 
mean. Now, a Gaussian probability distribution function has $0.000145$ of its area beyond the 
value $\pm 3.8$ standard deviations of the variable; this is obtained using the error 
function. This implies that half of that area, or $0.000073$, lies above and below $\pm 3.8$ 
standard deviations, respectively. This is consistent with the fact that for our sample size 
of $13454$ bins for each detector, each of $10$ s duration, one can expect one sample beyond  
$3.8$ standard deviations.  Note that the live time for each detector in Figure~\ref{fig2} is 
$13454 \times 10 = 134.54$ ksec and not the actual $\approx 139$ ksec because 
the {\it{extractor}} tool ignores bins of small fractional live time.

\begin{figure}[h]
\begin{center}
\includegraphics[keepaspectratio=true,scale=1.0,width=9.5cm]{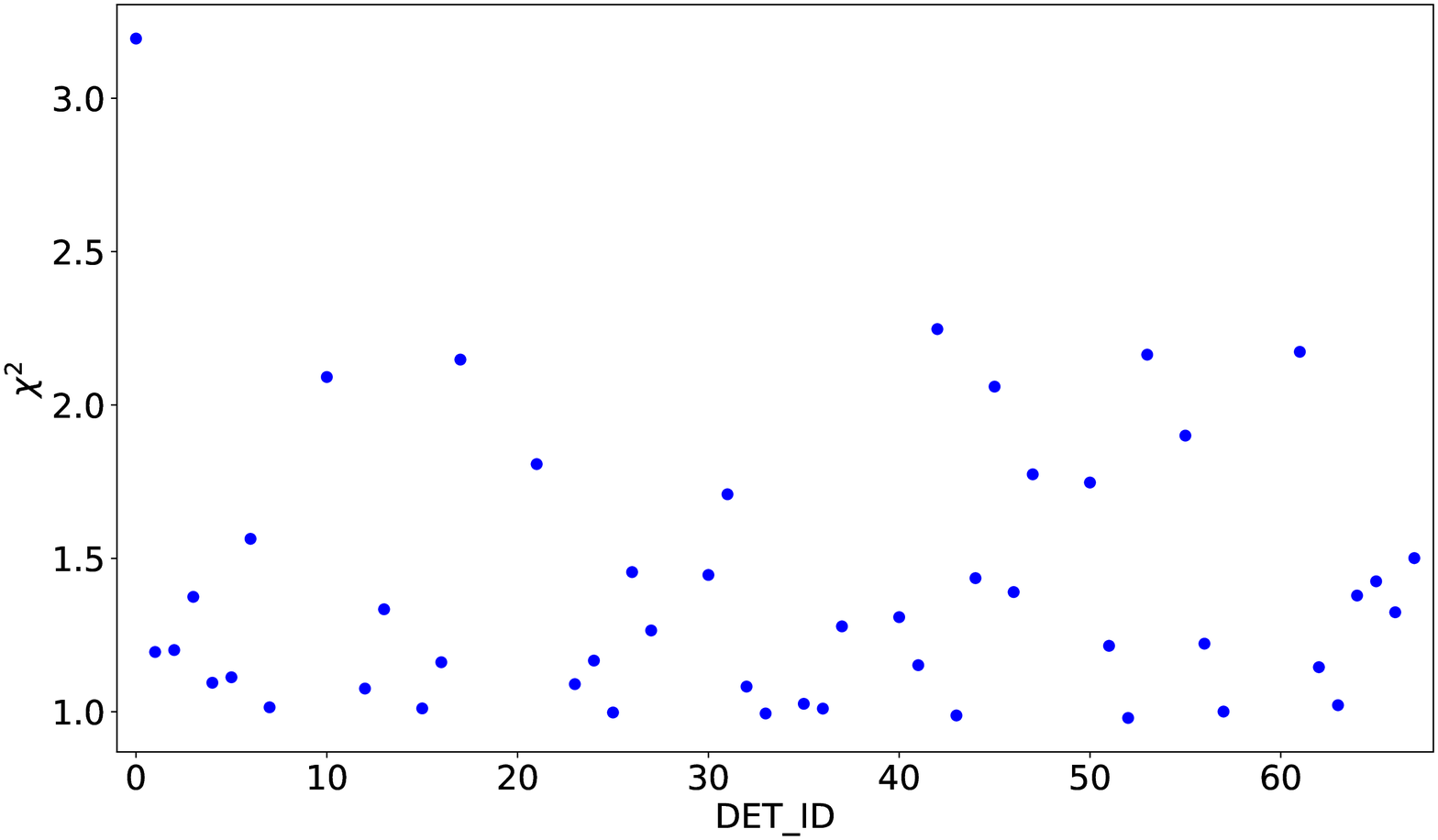}
\end{center}
\caption{
	Normalized $\chi^2$ between the LTLCs of the $49$ individual detectors and their 
	combined LTLC shown in Figure~\ref{fig1}.
        }
\label{fig3}
\end{figure}

Figure~\ref{fig3} compares the LTLCs of the $49$ individual detectors with that of all 
detectors combined (shown in Figure~\ref{fig1}), after normalizing each LTLC with its 
mean count rate. The $\chi^2$ per degree of freedom is below $1.5$ for $35$ detectors, 
between $1.5$ and $2$ for $7$ detectors and between $2$ and $2.25$ for $6$ detectors. 
The $\chi^2$ is derived using the formula $(y2 - y1)^2 / (y2 + y1)$, where $y1$ and $y2$ 
are the counts (not count rate) in the corresponding time bins of the two LTLCs being 
compared. Thus, only Poisson statistical fluctuations are being accounted for in 
Figure~\ref{fig3}. In such a complicated system such as {\it{NICER}}, it is reasonable to 
expect other kinds of fluctuations as well, not to mention minute source fluctuations 
also. Therefore, it is concluded that the LTLCs of all detectors are broadly similar.

{{
One is justified in normalizing each LTLC with its mean count rate, because the variations
of the latter in Figure~\ref{fig2} represent not merely statistical fluctuations, but
something more like, say, differences in effective area. The error bar on the mean counts 
in that figure is the standard deviation, and not the error on the mean; this was done to 
better compare the mean value with the maximum and minimum values. For 49 detectors 
there are $49 \times 48 / 2 = 1176$ distinct pairs, of which only $10$, $24$ and $43$ pairs 
have count rates differing by less than $3 \sigma$, $5 \sigma$ and $10 \sigma$, respectively. 
So the mean count rate of each detector is statistically distinct from that of another 
detector, and most probably represents the difference in their effective areas.
}}

\section{Raw spectrum}

\begin{figure}[h]
\begin{center}
\includegraphics[keepaspectratio=true,scale=1.0,width=9.5cm]{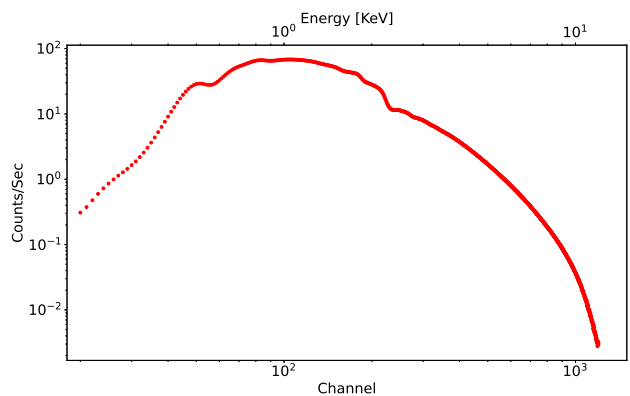}
\end{center}
\caption{
	Raw spectrum of the $49$ detectors combined, from channels $20$ to $1200$ ($0.2 - 12$ keV).
        }
\label{fig4}
\end{figure}

Figure~\ref{fig4} shows the combined raw spectrum (uncalibrated) of the $49$ detectors, 
consisting of $\approx 139$ ksec of data from $29$ ObsIDs, from energy $0.2 - 12$ 
keV. The sharp dip in spectrum at around $2.2$ keV is due to the Gold (Au) 
edge\footnote{https://heasarc.gsfc.nasa.gov/docs/heasarc/caldb/nicer/docs/xti/-NICER-Cal-Summit-ARF-2019.pdf};
the Gold is coated on the reflecting surface of the X-ray concentrators (XRC).

\begin{figure}[h]
\begin{center}
\includegraphics[keepaspectratio=true,scale=1.0,width=8.5cm]{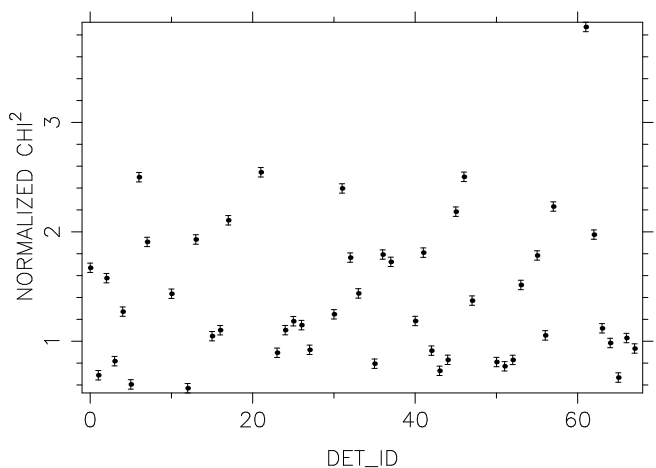}
\end{center}
\vskip+0.5cm
\caption{
	$\chi^2$ per degree of freedom between the raw spectrum of each detector and their
	combined spectrum in Figure~\ref{fig4}.
        }
\label{fig5}
\end{figure}

Figure~\ref{fig5} shows the comparison between Figure~\ref{fig4} and the corresponding spectrum
of each detector, in terms of the $\chi^2$ per degree of freedom obtained over $1100$ channels 
in the energy range $1 - 12$ keV, obtained using the same formula as in Figure~\ref{fig3}. 
{{ The lower energy range was ignored because here the $\chi^2$ was very high, mostly
due to the Carbon, Nitrogen and Oxygen edges at $\approx 0.3 - 0.5$ keV, and also due to the 
fact that our analysis is done with data beyond $1$ keV. }}
$41$ of the $49$ $\chi^2$ values are below $2.0$; the extreme value of $3.9$ belongs to detector 
$61$. Once again, one expects non-Poisson fluctuations to exist in the data. {{ The spread in 
$\chi^2$ is much larger than their formal statistical errors. However most of the $\chi^2$ gets
contribution from the $2.2$ keV Gold edge. Therefore, one does not expect better agreement 
than this. It is therefore concluded that the raw spectra of the $49$ detectors are statistically 
similar.

Figure~\ref{fig5} is not merely Figure~\ref{fig2} plotted in a different way; each is sensitive 
to different physical parameters. The latter is sensitive to time variations of effective area, 
deadtime, etc., while the former is sensitive to optical loading of the detectors, calibration
of the gain of the receivers that assigns an energy channel to each photon event, etc.
}}

\begin{figure}[h]
\begin{center}
\includegraphics[keepaspectratio=true,scale=1.0,width=8.5cm]{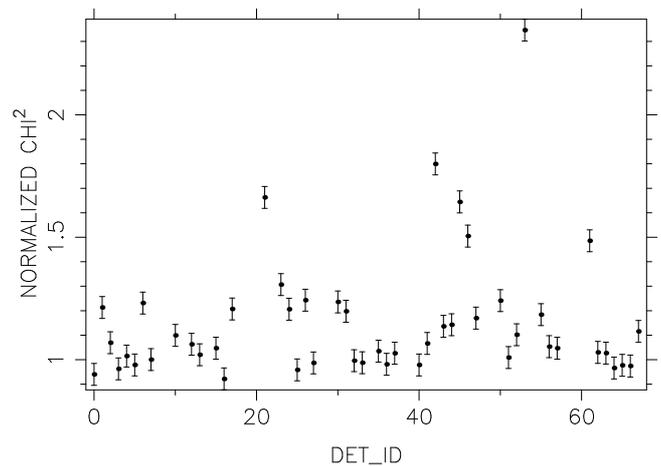}
\end{center}
\caption{
	Normalized $\chi^2$ between the raw spectra of earlier and later halves (durations) 
	of the data for each detector.
        }
\label{fig6}
\end{figure}

Figure~\ref{fig6} explores whether the spectrum of each detector changes as a function of time.
It displays the $\chi^2$ between data earlier to and later than the date $2018$ Nov $10$, which
roughly divides the data into two halves. The $\chi^2$ is estimated using the same formula as 
in Figure~\ref{fig3}, after scaling the two spectra with their respective live times, because
the cut-off date does not divide the data into exactly two equal halves. Just as in the previous 
figure, the $\chi^2$ values in Figure~\ref{fig6} are reasonable. It is therefore concluded that 
the raw spectrum of each detector does not change as a function of time during the $2.35$ year 
of observations.

{{
In both Figure~\ref{fig5} and Figure~\ref{fig6}, most of the estimated $\chi^2$ are 
larger than their formal errors. An extreme position to take would be to conclude
that no detector behaves like any other. However, the acceptance or rejection of a 
hypothesis based on $\chi^2$ depends upon the confidence level, which can be 
subjective in some contexts. In complicated systems such as {\it{NICER}} one expects
large  $\chi^2$ due to unknown or immeasurable factors. In principle, one can chose 
for further analysis only those detectors whose $\chi^2$ is less than, say, 2.0, and 
live with the corresponding reduced sensitivity. Here statistics can not help one -- 
only experience, intuition and the context will. To cite an example, in the good old
days radio astronomers considered the detection of a point source as positive if it 
was detected at a signal to noise ratio of $5$. One can ask why $5$, and not $6$ or 
$4$? The answer is that this number was arrived at through common experience and not
through pure statistics.
}}

\section{Deadtime}

\begin{figure}[b]
\centering
\includegraphics[width=8.5cm]{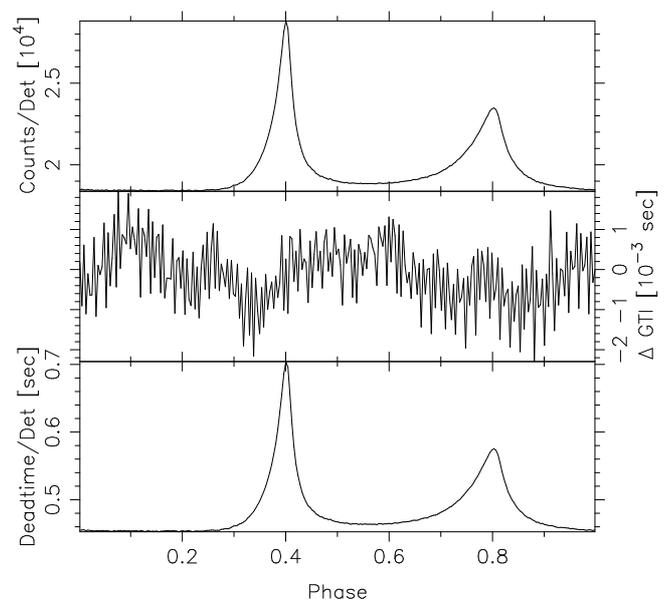}
\caption{
	Top to bottom panels show mean values of X-ray counts per detector, GTI difference, 
	and deadtime per detector using all counts, respectively, per phase bin for data of 
	ObsID $1013010147$, after applying the method of \cite{Wilson-Hodge2018}.
	{{The reference epoch of the first peak (phase 0.4) is MJD $58479.425994152342$.}}
        }
\label{fig7}
\end{figure}

\cite{Wilson-Hodge2018} describe the procedure for estimating the dead time for {\it{NICER}}. There 
are two sources of deadtime. First, after reception of an event, the detector is unable to 
process further events for some time, during which it processes the previous event. This
duration is recorded in the DEADTIME column of the event files. Second, if the source is 
bright, or the background is flaring, data buffers may saturate in the Measurement/Power Units 
(MPUs), which may stall recording of further events until the MPUs are reset. This causes loss 
of GTI, or equivalently, deadtime on the source. In this work the former will be called 
detector deadtime while the latter will be called GTI deadtime. The detector deadtime is 
contributed by all events, and not just the X-ray events. In \cite{Vivekanand2020} only the 
detector deadtime was taken into account and that too only for X-ray events.

The procedure of \cite{Wilson-Hodge2018} was repeated almost exactly for four ObsIDs -- 
$1013010113$ and $1013010147$, having the shortest ($120$ s) and longest ($23364$ s) live 
times respectively, and $1011010201$ and $1013010122$ having intermediate live times ($1841$ 
and $3850$ s respectively). While \cite{Wilson-Hodge2018} used $100$ phase bins per period,
and $1000$ phase bins for estimating GTI deadtime, the corresponding numbers here are $256$
and $1000$. The results are similar for all four ObsIds, so the detailed results for ObsID 
$1013010147$ alone will be presented here.

The top panel of Figure~\ref{fig7} shows the mean X-ray counts per detector as a function of 
phase for ObsID $1013010147$ (i.e., the FLC).  This has been obtained using cleaned events 
\citep{Wilson-Hodge2018}. The middle panel shows the mean GTI per phase bin after subtracting 
the expected mean value of $91.263$ s, which is obtained by dividing the live time by the 
number of phase bins. This has been obtained from the GTIs of the individual MPUs. The last 
panel shows the mean deadtime per detector, obtained using all events \citep{Wilson-Hodge2018}.
In each panel of Figure~\ref{fig7}, the results are first obtained per MPU and then averaged.

Two points are noteworthy in Figure~\ref{fig7}. First, in the middle panel, the difference 
in GTI per phase bin, between the observed and the expected values, has a mean value of 
$-0.1$ ms, and an rms of $0.8$ ms; the peak to peak variation is $4.25$ 
ms. The mean value of GTI differs from the expected value by $-0.0001 / 91.263 \approx
-0.0001$\%, while the peak to peak variation corresponds to $0.00425 / 91.263 \approx 
0.005$\% variation. This implies that the GTI per phase bin is almost 
constant across the FLC and that its observed value is very close to the expected value;
and so there is no significant loss of GTI due to MPU buffer overflows. This is further 
supported by the fact that, had there been GTI losses, it would have been anti-correlated 
with the FLC in the top panel which appears not to be the case.

Second, the detector deadtime fraction in the last panel of Figure~\ref{fig7} in the
off-pulse region (phase $< 0.2$) is $0.50$\%, while at the pulse peak it is $0.77$\%. The 
corresponding numbers in Figure~\ref{fig1} of \cite{Vivekanand2020} are $0.46$\% and 
$0.73$\%. Thus the detector deadtime fraction obtained using all events is only $\approx 
5 - 9$\% larger than that obtained using only X-ray events. However, these numbers are 
approximate. The correct method of estimating the deadtime fraction or count rate is to 
first estimate them per MPU, and then average \citep{Wilson-Hodge2018}, since the GTI 
per MPU as well as the number of detectors per MPU could differ across the MPUs.

\begin{figure}[h]
\centering
\includegraphics[width=8.5cm]{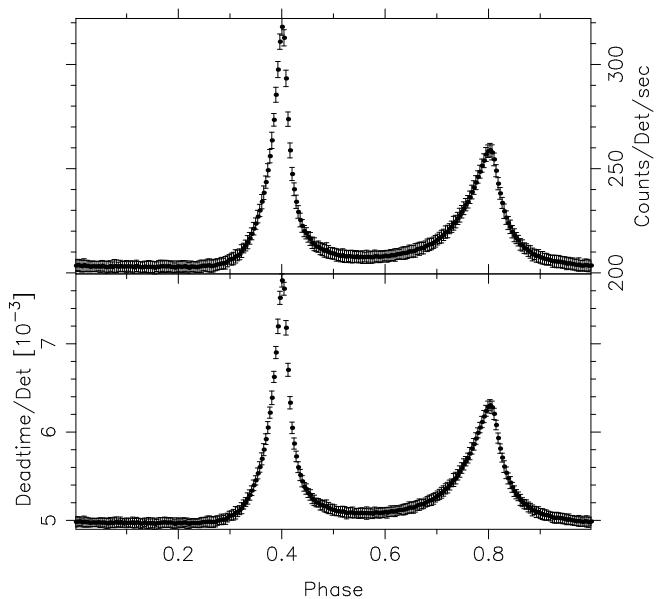}
\caption{
	Top and bottom panels show the mean values of count rate per detector {{(deadtime 
	corrected)}}, and deadtime fraction per detector, respectively, per phase bin for 
	data of ObsID $1013010147$, after applying the method of \cite{Wilson-Hodge2018}.
	{{The reference epoch of the first peak (phase 0.4) is MJD $58479.425994152342$.}}
        }
\label{fig8}
\end{figure}

This is done in Figure~\ref{fig8}. The mean off pulse deadtime fraction is $0.497(1)$\%, 
while that at the pulse peak is $0.772(7)$\%; these numbers are very similar to the 
approximate values above. 

The results of Figures $7$ and $8$ hold for data of the other three ObsIDs also, except
for lower statistical significance on account of lower live times.

Therefore, if the GTI deadtime is non existent for the Crab pulsar, the difference 
between detector deadtime obtained using all events and that obtained using only 
X-ray events, must depend upon the difference in the number of these two event
types. This is demonstrated in Table~\ref{tbl1}, which contains event statistics 
from the filter file of that ObsID; for example the filter file for ObsID 
$1013010113$ would be ``ni$1013010113$.mkf'' found in the directory AUXIL. The 
filter file contains the numbers of various kinds of events registered by {\it{NICER}}, 
logged once every second in time. In particular, it contains two columns labeled 
TOT\_ALL\_CNTS and TOT\_XRAY\_CNTS. The former contains the number of all types of
events, while the latter contains the number of what are deemed to be X-ray events,
but are not guaranteed to be so\footnote{https://heasarc.gsfc.nasa.gov/docs/nicer/mission\_guide/}.
The second and third columns of Table~\ref{tbl1} contain the sum of TOT\_ALL\_CNTS 
and TOT\_XRAY\_CNTS, respectively, for the duration within the GTI of the 
corresponding event file, for $49$ useful detectors. For example, the number of 
TOT\_ALL\_CNTS and TOT\_XRAY\_CNTS for ObsID $1013010113$ registered within the 
GTI are actually $1114762$ and $1094359$, respectively, for $52$ detectors; after 
deleting the contribution of detectors $14$, $34$ and $54$, found in the columns 
MPU\_ALL\_COUNT and MPU\_XRAY\_COUNT in the filter file, the numbers become 
$1049364$ and $1030023$, respectively.

The fourth column of Table~\ref{tbl1} gives the {{ ratio of the number of events of 
column $3$ that were eventually classified as X-ray events, obtained in the keyword
NAXIS2 in the corresponding event file, and TOT\_ALL\_CNTS. }}
Table~\ref{tbl1} has been made for all $29$ ObsIDs used in this work, and the mean 
value of the above ratio is $0.922(5)$. This demonstrates that the difference between 
actual X-ray counts and all counts for the Crab pulsar data of {\it{NICER}} is typically 
about $7.8(5)$\%{{, which is a small number.}}

There are two conclusions so far in this section. First, The difference between the
number of all events and the number of X-ray events for the {\it{NICER}} data of the Crab 
pulsar is quite small ($\approx 8$\%). This is not surprising, since the event 
rate for the Crab pulsar is dominated by events from the source; for much less 
luminous sources, the event rate would be dominated by particle events, etc. 
Second, the GTI deadtime is negligible for the Crab data of {\it{NICER}}. This is also not 
surprising, since source count rates of about $18846$ per second (for 49 detectors) 
begin to show such 
effects\footnote{https://heasarc.gsfc.nasa.gov/docs/nicer/data\_analysis/-nicer\_analysis\_tips.html\#Deadtime}, 
and the Crab pulsar's peak count rate is only about ${{\approx 17000}}$ counts per second. Further,
MPU buffer overflow depends upon the product of the count rate and the duration for
which it is maintained; 

\onecolumn

\begin{table}[t]
\begin{center}
\caption{
	Statistics of events in the four filter (MKF) files. The four columns contain the 
	ObsID, total counts, presumed X-ray counts and {{the ratio of actual X-ray counts
	and total counts}}, for $49$ useful detectors.
        } \label{tbl1}
\begin{tabular}{|c|c|c|l|}
\hline
ObsID        & TOT\_ALL\_CNT & TOT\_XRAY\_CNT & NAXIS2 / TOT\_ALL\_CNT     \\
\hline
$1013010113$ & $1049364$      &  $1030023$      & $0.926$ \\
\hline
$1011010201$ & $13865806$     & $13626936$      & $0.931$ \\
\hline
$1013010122$ & $34387222$     & $33712711$      & $0.926$ \\
\hline
$1013010147$ & $202406166$    & $197369215$     & $0.919$ \\
\hline
\hline
\end{tabular}
\end{center}
\end{table}

\twocolumn

\begin{figure}[h]
\centering
\includegraphics[width=8.5cm]{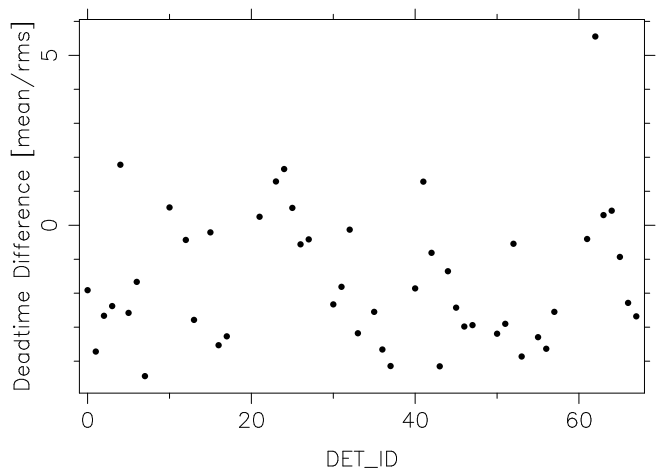}
\caption{
	Comparison of detector deadtime fractions of the $49$ detectors.
        }
\label{fig9}
\end{figure}

\noindent 
for the Crab pulsar the peak count rate is maintained only for a few ms, so the probability 
of buffer overflow is small. Furthermore, the stringent filtering of this work might have 
ensured that high particle durations are excluded, which reduces the probability of buffer 
overflow in the remaining data.

Figure~\ref{fig9} compares the detector deadtime fraction of each of the $49$ detectors 
with the combined detector deadtime fraction of all $49$ detectors, across $256$ phase bins
in a period. It estimates the $256$ differences of the two curves and computes their mean 
and standard deviation. Ideally, the mean should be zero for each detector and the standard
deviation should reflect the average measurement error of the detector deadtime fraction
across a period. Figure~\ref{fig9} shows the ratio of mean and standard deviation for each
detector. The mean value of this ratio in Figure~\ref{fig9} is $-1.6 \pm 2.0$. If one 
ignores the extreme value of $5.6$ of detector $62$, the results are $-1.7 \pm 1.7$. Thus
the mean value of this ratio is consistent with being zero. 

In Figure~\ref{fig9}, the relatively large values of standard deviations {{(not standard 
deviation of the ratio mean/rms)}} for each detector 
are due to two factors -- the reduced number of photons in each phase bin for a single detector 
and, to a lesser extent, the intrinsic distribution of detector deadtimes in the column DEADTIME 
in the event files. It should be recalled that the quantity being estimated here (detector 
deadtime fraction) is very small, of the order of $0.5$\% to $0.8$\%; the difference of two such 
quantities is typically of the order of $\approx 0.05$\% to $0.005$\%.  Now, the mean number of 
events in a phase bin for a single detector is $1570142182 / 49 / 256 \approx 125171$; the 
Poisson fluctuation in this quantity is of the order of $\sqrt(125171) / 125171 \approx 0.28$\%; 
over $256$ phase bins the accuracy of the estimated mean (due to the Poisson process alone) would 
be $0.28 / 16 \approx 0.02$\%, which is comparable to the quantity being estimated. Similarly,
the mean detector deadtime for ObsID $1013010147$ is $23.1 \pm 4.2$ $\mu$s \citep{Vivekanand2020}.
The fractional error is $4.2 / 23.1 \approx 18.2$\%. The mean detector deadtime estimated using
$125171$ events will have an accuracy of $18.2 / \sqrt(125171) \approx 0.05$\%, which is also
comparable to the quantity being estimated.

\begin{figure}[h]
\centering
\includegraphics[width=8.5cm]{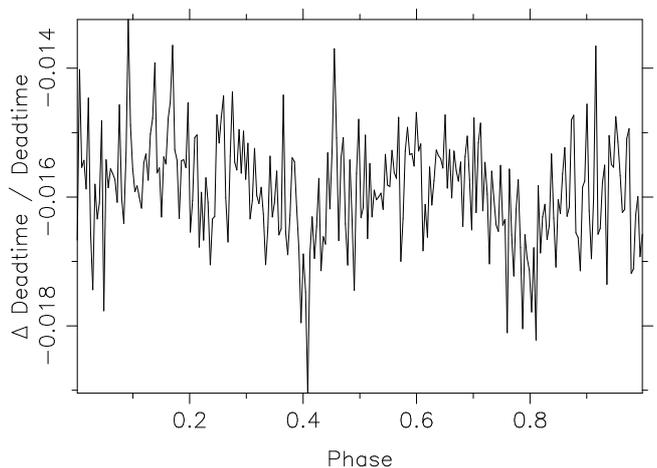}
\caption{
	Fractional difference in detector deadtime fraction between the first and
	second halves (durations) of the data. {{The reference epoch of phase 
	0.4 is MJD $58479.425994152342$.}}
        }
\label{fig10}
\end{figure}

The time evolution of detector deadtime is studied by estimating the deadtimes of the 
first and second halves of the data, for all $49$ detectors combined, with the cutoff 
date of $2018$ Nov $10$ that was used in Figure~\ref{fig6}. The quantity plotted in 
Figure~\ref{fig10} is $(z2 - z1) / (z2 + z1) * 2.0$ as a function of phase, where $z1$ 
and $z2$ are the detector deadtime fractions in the first and second halves of the data, 
respectively. The mean value of the fractional difference is $-0.016(1)$; i.e.,
the earlier and later detector deadtime fractions differ by about $1.6(1)$\%. It
is therefore concluded that there is no evolution of detector deadtimes as a function 
of epoch, during the $2.35$ years of observations studied here.

\begin{figure}[h]
\centering
\includegraphics[width=8.5cm]{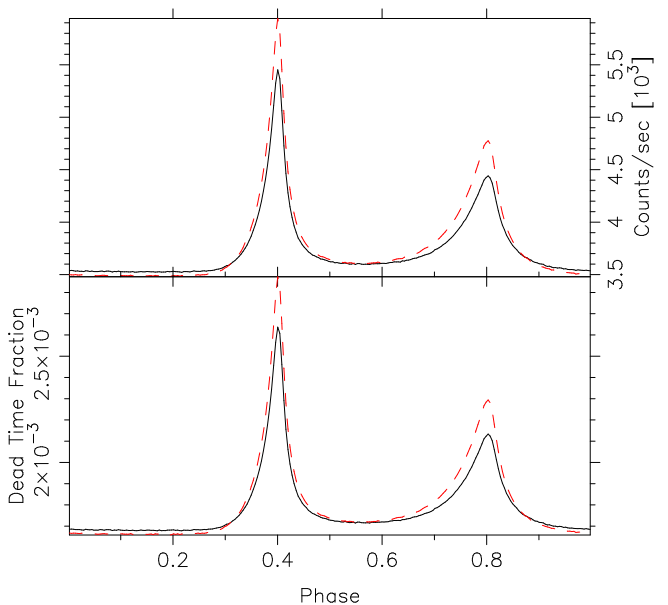}
\caption{
	Top panel: FLC for data of all $49$ detectors in the energy range $1.0 - 1.62$ 
	keV (dashed line), and in the energy range $1.62 - 10.0$ keV (solid line). Bottom 
	panel: The deadtime fraction of the corresponding FLC in the top panel.
	{{The reference epoch of the first peak (phase 0.4) is MJD $58479.425994152342$.}}
        }
\label{fig11}
\end{figure}

Finally, Figure~\ref{fig11} investigates the dependence of detector deadtime upon photon energy. 
The combined data of all $49$ detectors was divided roughly into two equal parts -- one consisting
of photons with energy between $1.0 - 1.62$ keV, and the other between $1.62 - 10.0$ keV. The FLC 
and deadtime fraction for each energy band is shown in the top and bottom panels of 
Figure~\ref{fig11}, respectively. The spectral evolution of the FLC with energy is evident in the 
top panel of Figure~\ref{fig11}. The deadtime fraction curves mirror this evolution in the bottom 
panel, so it is not possible to directly compare the two curves. However, if the detector deadtime 
in a  phase bin depends only upon the photon count rate in that bin and not upon the photon's 
energy, then the deadtime fraction would have the same linear dependence upon the count rate for 
the two energy bands. So a linear fit was done to the deadtime fraction against the count rate in 
each energy band (see \cite{Vivekanand2020} for justification of linear dependence), and the 
results are given in Table~\ref{tbl2}.

\begin{table}[h]
\begin{center}
\caption{
	Results of straight line fit to the data of detector deadtime and photon count rate in 
	Figure~\ref{fig11} -- intercept $A$, slope $B$ and the standard deviation of the data 
	with respect to the model $C$.
        } \label{tbl2}
\begin{tabular}{|c|c|c|c|}
\hline
Energy Band      & $A$ ($\times 10^{-5})$    & $B$ ($10^{-7}$)  &  $C$ ($10^{-6}$) \\
\hline
$1.0 - 1.62$ keV & $-7.0(2)$               & $4.961(5)$    & $2.78$ \\
\hline
$1.62 - 10$ keV  & $-5.6(2)$               & $4.924(7)$    & $3.61$ \\
\hline
\hline
\hline
\end{tabular}
\end{center}
\end{table}

The two slopes are almost the same, while the intercepts are negligible. This proves that the
detector deadtime fraction depends only upon the count rate and not upon the photon energy.

\section{Calibrated phase averaged spectrum (PAS)}

\begin{figure}[h]
\centering
\includegraphics[width=8.5cm]{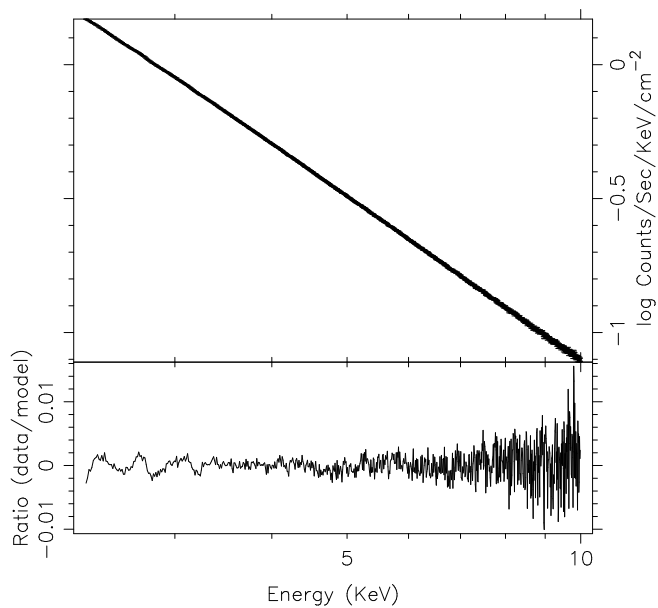}
\caption{
	Top panel: Phase averaged spectrum of the Crab {{nebula plus}} pulsar. Bottom panel: ratio of data
	and model. {{ Channel dependent effective area normalization is done using the 
	"{\it{setplot area}}" command of XSPEC.  }}
        }
\label{fig12}
\end{figure}

The top panel of Figure~\ref{fig12} shows the calibrated phase averaged spectrum (PAS) of 
the Crab {{nebula plus}} pulsar using all $49$ useful detectors, for the $29$ ObsIDs with a live time of 
$\approx 139$ ksec. The spectrum was obtained using the {\it{extractor}} tool. It was 
analyzed using XSPEC\footnote{https://heasarc.gsfc.nasa.gov/docs/xanadu/xspec/manual/manual.html} 
version $12.11.0$, using {\it{nixti20170601\_combined\_v002.rmf}} as the main response file and 
{\it{nixtionaxis20170601\_combined\_v004.arf}} as the ancillary response file. These differ
from the corresponding files found in CALDB version XTI($20200722$), viz. 
{\it{nixtiref20170601v002.rmf}} and {\it{nixtiaveonaxis20170601v004.arf}}, in that the latter 
two are relevant for observations using all $52$ detectors of {\it{NICER}}, while the former two are 
relevant for observations using only $49$ detectors; see the relevant analysis 
thread\footnote{https://heasarc.gsfc.nasa.gov/docs/nicer/analysis\_threads/arf-rmf/} at the 
{\it{NICER}} site. Channels $230$ to $1000$ were used (energy range $2.3 - 10$ keV). No background 
spectrum was used since that becomes important for the Crab pulsar only at much higher energies 
\citep{Madsen2015}{{, as mentioned in section $2$. \cite{Ludlam2018} state that the background
fields have count rates of $0.7 - 2.3$ counts/s, while the Crab pulsar's average count rate is
$\approx 11020$ counts/s, for $52$ detectors.}} 
The default photo ionization cross-sections and solar abundances were used ({\it{vern}} and {\it{angr}}). 
An absorbed power law model was fit to the data of Figure~\ref{fig12}; the 
parameters of the model are the column density of absorbing Hydrogen $N_H$, the power law 
index $\Gamma$, and the power law normalization $A$. The results are: $N_H = 0.210(3) 
\times 10^{22}$ cm$^{-2}$, $\Gamma = 2.050(1)$; and $A = 8.82(1)$; the $\chi^2$ per degree 
of freedom is $1.99$, for $769$ degrees of freedom. The errors are obtained using the {\it{error}} function and represent 
$90$\% confidence levels; at higher decimal places the lower and upper limits of the errors 
are asymmetric. 

These results are consistent with the results of {\it{XMM-Newton}} in the $0.6 - 9$ keV
band for the overall Crab region \citep{Kirsch2006}, in their Table $2$ and Figure $6$. 
Their values are $N_H = 0.260(1) \times 10^{22}$ cm$^{-2}$ and $\Gamma = 2.046(3)$, 
with $\chi^2 = 1.99$ per degree of freedom.

These results also compare well with the results of {\it{NuSTAR}} \citep{Madsen2015} in their
Figure $1$, in the energy range $3 - 78$ keV. They fixed the value of $N_H$ at 
$0.2 \times 10^{22}$, since their data is not sensitive to this parameter, and use
the {\it{wilms}} solar abundances. Their $\Gamma = 2.0963(4)$, and they do not mention
the $\chi^2$. Their $\Gamma$ is in the same numerical range as our result considering 
the fact that their energy range is much larger.

The bottom panel of Figure~\ref{fig12} shows the ratio of the data to the absorbed power 
law model curve. As found in Figure~\ref{fig1} of \cite{Madsen2015}, the variations are 
of the order of $1$\%, presumably indicating the accuracy of the spectral calibration of 
these instruments.

As mentioned by \cite{Madsen2015}, the non piled up instruments covering the $1 - 100$ 
keV band agree on a power law index of $2.10(2)$ for the PAS of the Crab nebula plus pulsar, 
without any break or curvature in this band; see also \cite{Kirsch2005}. \cite{Kouzu2013}
also confirm this result using the HXD detector aboard {\it{Suzaku}}.  Our result is 
consistent with this statement. Moreover, the more important result is the variation of 
$\Gamma$ across the rotation phase, and not its absolute value, since that is correlated 
with $N_H$ (see Figure $2$ of Massaro et al. 2000; see also XSPEC manual).

It is therefore concluded that the PAS of the Crab pulsar and nebula
in Figure~\ref{fig12} is consistent with earlier studies.

\begin{figure}[h]
\centering
\includegraphics[width=8.5cm]{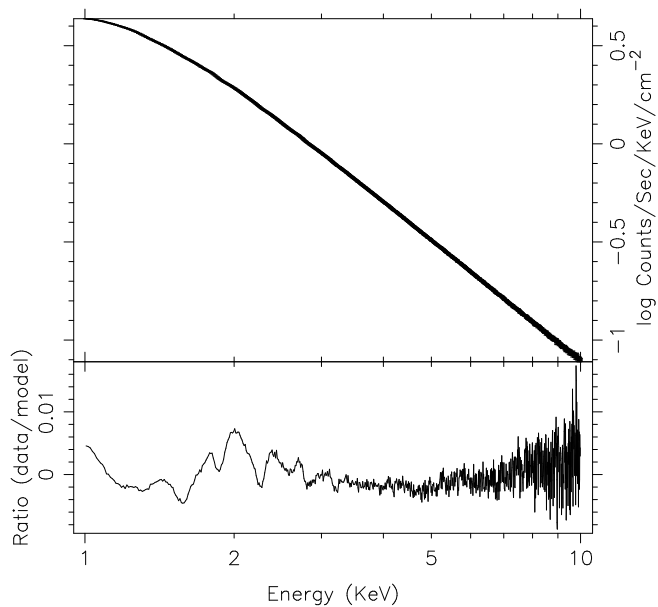}
\caption{
	The same as in Figure~\ref{fig12}, with a wider energy range.
        }
\label{fig13}
\end{figure}

Figure~\ref{fig13} is similar to Figure~\ref{fig12}, except that lower energy limit is
$1.0$ keV. The ratio of the data to the model in its lower panel has variations of the
order of $1$\%, the same as in the latter figure. The fit parameter values are: $N_H = 
0.2943(1) \times 10^{22}$ cm$^{-2}$, $\Gamma = 2.0651(2)$; and $A = 9.095(2)$. These
are consistent with the values obtained above for Figure~\ref{fig12}, but the errors
are unreliably smaller, because the $\chi^2$ per degree of freedom is $44.4$, 
contributed mainly by the channels between energies $1.0 - 2.3$ keV. This 
appears to be due to the very high photon counts in the lower energy channels. This
problem disappears when the PRS is computed in the next section,
since the number of photons in each energy channel is now less by the number of phases
in the period.

\section{Calibrated phase resolved spectrum (PRS)}

The PRS of the Crab pulsar has been studied over the last two and half decades by several 
workers at both the soft and the hard X-ray energies, using different X-ray observatories 
({\it{RXTE}}, {\it{BeppoSAX}}, {\it{XMM-Newton}}, {\it{Chandra}}, {\it{NuSTAR}}  and {\it{Insight-HXMT}}) \citep{Pravdo1997, Massaro2000, 
Massaro2006, Kirsch2006, Weisskopf2011, Ge2012, Madsen2015, Tuo2019}. The highest phase 
resolution achieved so far has been typically $\approx 1/100$ in phase; even though {\it{BeppoSAX}} 
\citep{Massaro2000, Massaro2006} and {\it{RXTE}} \citep{Ge2012} start with $300$ and $1000$ phase 
bins for the FLC, respectively, they have had to estimate the PRS over $\approx 100$ phase 
bins to maintain statistical significance. The following sub section presents the PRS of the
Crab pulsar obtained by {\it{NICER}} over $128$ phase bins. The next sub section obtains the PRS
over $1024$ bins, although the number of independent phase bins is probably $512$ or better.

The $29$ ObsIds used in this work already have their $\nu$ and $\dot \nu$ (each at its
reference epoch) estimated by \cite{Vivekanand2020}, along with their formal errors. These 
were verified in those cases where the event files had sufficient number of photons in spite 
of the stringent filtering of this work. For each ObsID, using its start and stop times 
available in the corresponding event file, a phase error was calculated, that would result 
if the $\nu$ and $\dot \nu$ used were different by twice their formal errors. Only those 
ObsIDs were retained that had the phase error less than $1/2048$. This eliminated the ObsIDs 
$1011010201$ and $1013010119$, leaving $27$ ObsIDs for further analysis. Next, FLCs were 
formed for each of the $27$ ObsID in $1024$ phase bins. These were then cross correlated 
with the FLC of ObsID $1013010147$, after shifting it in phase so that the first peak of 
its FLC lies at phase $0.4$. The $26$ phase offsets thus obtained had a typical error of 
$\approx 10^{-4}$ or better. These phase offsets were then used along with the corresponding 
$\nu$ and $\dot \nu$ to compute the phase of each photon event in the $27$ ObsIDs (see 
\cite{Vivekanand2020} for details).

\subsection{Low phase resolution PRS}

\begin{figure}[t]
\centering
\includegraphics[width=8.5cm]{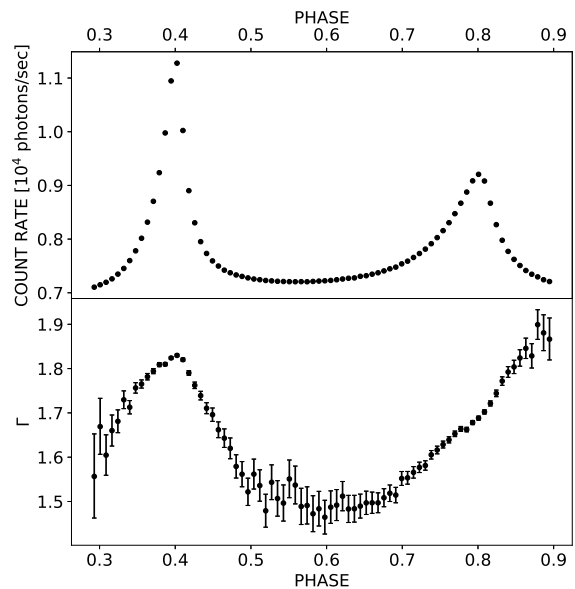}
\caption{
	Top panel: FLC of the Crab pulsar at a resolution of $1/128$ in rotation phase,
	in the energy range  $1 - 10$ keV. Bottom panel: Power law index $\Gamma$ of 
	the PRS of the Crab pulsar at the same resolution, over the same energy range. 
	The background spectrum is estimated from data in the phase range $0.0$ to $0.2$.
	{{The reference epoch of the first peak (phase 0.4) is MJD $58479.425994152342$.}}
        }
\label{fig14}
\end{figure}

Figure~\ref{fig14} shows the calibrated PRS of the Crab pulsar using all $49$ useful detectors 
($27$ ObsIDs, live time $\approx 136$ ksec), over the energy range $1 - 10$ keV.
The spectra were obtained with $128$ phase bins per cycle, using the {\it{extractor}} tool, 
and were analyzed using XSPEC. The background spectrum was extracted from the phase range $0.0$ 
to $0.2$. The response and ancillary response files used are those mentioned earlier. Spectral 
channels below $100$ and above $1000$ were ignored. The default photo ionization cross-sections 
and solar abundances were used. Details are given in Appendix C.

It is known that, while the PAS of the Crab pulsar fits well to a single power law model over 
the energy range $\approx 1 - 100$ keV, the PRS requires either a break in power 
law at around $11 - 13$ keV \citep{Madsen2015}, or a curved spectrum \citep{Massaro2000, 
Massaro2006}. The {\it{NICER}} energy range justifies the use of a single power law. So in 
Figure~\ref{fig14} a single absorbed power law was fit to each of the $78$ phase resolved 
spectra, The phase ranges $0.2 - 0.3$ and $0.9 - 1.0$  do not have sufficient number of 
photons for a reasonable fit, after background subtraction.

For the fit, the spectra at all phases must have the same absorbing column density $N_H$. As 
mentioned earlier, the power law index $\Gamma$ is correlated with $N_H$; and the absolute 
value of $\Gamma$ is not as important as its variation with rotation phase. Therefore $N_H$ was 
fixed at the value $0.36 \times 10^{22}$ cm$^{-2}$ in Figure~\ref{fig14}  (Massaro et al. 2006;
see also Table $2$ of Weisskopf et al. 2011, and references in Ge et al. 2012). As before, the 
errors in Figure~\ref{fig14} represent the $90$\% confidence level.

The qualitative variation of $\Gamma$ with phase in Figure~\ref{fig14} is consistent with the 
results obtained with the {\it{BeppoSAX}} detector MECS, in the energy range $1.6 - 10$ keV 
(Figure $3$b of Massaro et al. 2000 and Figure $1$ of Massaro et al. 2006). It is also consistent
with the results of {\it{RXTE}} in the energy range $5 - 60$ keV (Figure $1$ of Pravdo et al. 1997, 
and Figure $12$ of Ge et al. 2012). However, the {\it{RXTE}} results are obtained by fitting only a 
single power law over the much larger energy range. It is also consistent with the results of
{\it{XMM-Newton}} in the energy range $0.6 - 6.5$ keV, although their phase resolution is relatively 
poor (see Figure $11$ of Kirsch et al. 2006). It is also consistent with the results of {\it{NuSTAR}}, 
who fit a broken power law in the energy range $3 - 78$ keV (bottom panel of Figure $3$ of 
\cite{Madsen2015}). It appears to also be consistent with the results of Insight-HXMT, although 
their energy range is much larger, $11 - 250$ keV, and they appear to fit a single power law 
in this entire energy range (Table $3$ and Figure $5$ of \cite{Tuo2019}).  Chandra's energy 
range was much smaller, $0.3 - 3.8$ keV, and it had relatively much fewer photons; so 
comparison with this work is difficult.

It is therefore concluded that the results of Figure~\ref{fig14} are qualitatively consistent 
with previous results. Next some quantitative comparisons will be made.

The PRS of {\it{NuSTAR}} \citep{Madsen2015} that can be compared with this work is given in the 
$50''$ extraction region section of their Table $2$. The peak of their FLC appears to be 
at phase $0.1$ in the bottom panel of their Figure $3$, which corresponds to phase $0.4$ in 
Figure~\ref{fig14} above. So their three bins straddling the peak, with phase boundaries 
$0.0 - 0.07$, $0.07 - 0.14$ and $0.14 - 0.21$, correspond to the phase boundaries $0.30 - 
0.37$, $0.37 - 0.44$ and $0.44 - 0.51$, respectively in Figure~\ref{fig14} above. These three 
bins were chosen for comparison since they contain the highest number of photons in the PRS. 
Their background spectrum is extracted from their phase range $0.63 - 0.90$, which 
corresponds to the phase range $0.93 - 0.20$ in Figure~\ref{fig14} above. \cite{Madsen2015} 
fix the value of $N_H$ at $0.2 \times 10^{22}$ and use the {\it{wilm}} photo ionization 
cross-sections and {\it{vern}} solar abundances. 

\begin{table}[h]
\begin{center}
\caption{
	Comparison of the results of {\it{NuSTAR}} and the corresponding results of {\it{NICER}}.
	{{ Phase $0.1$ of the former corresponds to phase $0.4$ of the latter, which
	corresponds to the epoch MJD $58479.425994152342$.}}
        } \label{tbl3}
\begin{tabular}{|c|c|c|c|}
\hline
PHASE           & $\Gamma_1$ ({\it{NuSTAR}}) & $\Gamma$ ({\it{NICER}})  &  diff ($\sigma$) \\
\hline
$0.00 - 0.07$   & $1.66(4)$           & $1.72(2)$         & $-1.3$ \\
\hline
$0.07 - 0.14$   & $1.83(1)$           & $1.788(6)$        & $+3.6$ \\
\hline
$0.14 - 0.21$   & $1.55(4)$           & $1.60(3)$         & $-1.0$ \\
\hline
\hline
\hline
\end{tabular}
\end{center}
\end{table}

Doing the same with the {\it{NICER}} data in the corresponding phase bins, one obtains $\Gamma$ which 
is compared with their $\Gamma_1$ in Table~\ref{tbl2}, which is the first of the two spectral 
indices for a broken power law model that they use over the $3 - 78$ keV range, the break 
occurring at $13.1(4)$ keV.  The $\Gamma_1$ of \cite{Madsen2015} differs from the corresponding 
$\Gamma$ of this work by $-1.3$, $+3.6$ and $-1.0$ standard deviations, in the three phase bins;
the corresponding reduced $\chi^2$ of the fits are $0.99$, $1.04$ and $0.88$, respectively. It 
is therefore concluded that there is reasonable quantitative agreement between the results of 
{\it{NuSTAR}} and that of this work. {{ There are three aspects of {\it{NuSTAR}} data analysis 
which might explain the $+3.6 \sigma$ difference. Firstly, \cite{Madsen2015} do not specify where
exactly the peak of the main pulse lies -- I had to assume that it is at phase $0.1$ in their 
Figure $3$ by visual inspection. Any small difference in this value is likely to mix up the 
spectrum from the adjacent bins. Secondly, {\it{NuSTAR}} data is very severely affected by 
deadtime, as seen in their Figure $3$. If  \cite{Madsen2015} have used this for correcting the
spectrum also, then the effect would be felt most in the bin at the peak. Lastly, {\it{NuSTAR}}
is an imaging instrument, and the data presented in their Figure $3$ has been extracted from 
region between 17$^{''}$ and 50$^{''}$; a slightly different region might have yielded slightly
different result.
}}

\cite{Massaro2000} obtain the PRS of the Crab pulsar using the MECS instrument on {\it{BeppoSAX}}
at the first peak, in the phase interval $0.02667$ straddling the peak. They use the phase 
range of $0.60 - 0.83$ in their Figure $1$ to obtain the background spectrum, which 
translates to the phase range $0.0 - 0.23$ in Figure~\ref{fig14} above. They use $N_H = 
0.323 \times 10^{22}$ cm$^{-2}$. They obtain $\Gamma \approx 1.83$ in their Figure $5$ 
(numerical value is not given). Using the same parameters we obtain $\Gamma = 1.818(3)$,
which is a reasonable agreement.

Figure~\ref{fig14} was also produced using the energy ranges $2 - 10$ keV and $3 - 10$ keV, 
to check if the choice of the start of the energy range made any difference to the results. 
The $\Gamma$ variation was similar, except for reduced statistical significance on account 
of reduced number of photons. Two other energy ranges were also tried, viz.  $1 - 9$ keV 
and $1 - 8$ keV, but here the differences were imperceptible, as expected.

It is therefore concluded that the results of Figure~\ref{fig14} are quantitatively 
consistent with the results of {\it{NuSTAR}} and {\it{BeppoSAX}}.

The results of this and the following sub-section were obtained using both pairs of
{\it{rmf}} and {\it{arf}} calibration files; the difference between the two sets of
results is insignificant, except for the power law normalization coefficient $A$,
which differs by about $\approx 6$\% as expected, since that is the difference in
the effective area between the two calibrations.

\subsection{High phase resolution PRS}

\begin{figure}[h]
\begin{center}
\includegraphics[keepaspectratio=true,scale=1.0,width=9.5cm]{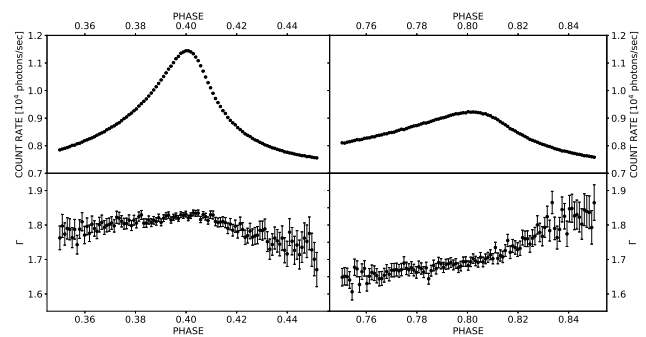}
\end{center}
\caption{
	The same as in Figure~\ref{fig14}, but at a phase resolution of $1/1024$. The left
	and right panels correspond to the data of the two peaks of the FLC of the Crab
	pulsar.
	{{The reference epoch of the first peak (phase 0.4) is MJD $58479.425994152342$.}}
        }
\label{fig15}
\end{figure}

Figure~\ref{fig15} shows the equivalent of Figure~\ref{fig14} at the two peaks of the FLC
of the Crab pulsar, at a phase resolution of $1/1024$. The analysis method is the same as in
Figure~\ref{fig14}, except that due to further reduction in photon counts per phase bin, 
energy bins that had less than $4$ photons were ignored in each PRS.

\cite{Ge2012} display high resolution FLC at the two peaks, but in different energy bands,
but they do not display $\Gamma$ at  this high resolution.

{{
In the top left panel of Figure~\ref{fig15} the count rate rises from $9334.3$ at phase 
$0.38$ to $11442.1$ at phase $0.40$, then falls to $8758.6$ at phase $0.42$, indicating
that at the first peak, the FLC of the Crab pulsar rises slowly and falls faster. The
intermediate count rates of $10383.3$ and $10101.4$ at phases $0.39$ and $0.41$,
respectively, confirm this trend.

In the top right panel of Figure~\ref{fig15} the count rates are $8728.8$, $8999.4$, 
$9232.1$, $9022.7$, and $8473.4$ at phases $0.78$, $0.79$, $0.80$, $0.81$ and  $0.82$,
respectively, not showing the same clear trend.
}}

\section{Discussion}
{{

This section will begin with the scientific motivation for obtaining high resolution FLC and
PRS of the Crab pulsar at high energies. It was mentioned in section $1$ that these are
required to understand the high energy emission mechanism of RPPs. It was also mentioned that
the analysis of the PRS lagged far behind that of the FLC. One reason was probably the success 
obtained in theoretically producing the double peaked FLCs of several RPPs based only on the 
geometry of the magnetic field. The other reason {{was the}} relative difficulty in 
applying the basic radiative processes in a realistic magnetosphere of the RPPs, viz., 
curvature radiation (CR), inverse Compton scattering (ICS) and synchrotron radiation (SR) 
\citep{Cheng1986}. The availability of high resolution PRS, and its use in conjunction with
high resolution FLC, necessitates the invocation of these basic radiative processes in 
addition to magnetic field line geometry and magnetospheric processes. Such a problem is very 
complicated. In this work, the focus is restricted to the formation and properties of caustics 
in FLCs using the main results of this paper (Figure~\ref{fig15}).

\subsection{Two caustics or one?}

The left and right sections of Figure~\ref{fig15} show the FLC (top panel) and PRS (bottom 
panel) of the Crab pulsar at its two peaks. Now, it is generally believed that the FLC of 
the first peak (left section) is a caustic, due to its narrow and luminous peak. A 
legitimate query is whether the much wider and less luminous FLC of the second peak 
(right section) is not a caustic, or whether it was a caustic that was smoothed by some 
physical process. In the latter case, it would be reasonable to assume that the pre-widened 
caustic at the second peak had a PRS similar to that at the first peak. Then the physical 
process widening the second peak's FLC would also smooth its PRS. The challenge is to
come up with a physical process that can reshape the FLC at the first peak into that at the 
second peak, while simultaneously transforming the PRS at the first peak into that at the 
second peak.

At first glance this appears to be difficult. By deconvolving the FLC of the second peak
with the FLC of the first peak, it is possible to come up with a smoothing function that 
can transform the first FLC into the second. Naively this should also widen the PRS of the 
first peak, which would not look anything like the monotonically increasing function of 
phase in the second peak of Figure~\ref{fig15}. So one has to invoke a functional relation 
between the FLC smoothing function and the corresponding PRS smoothing function, to 
explain the data in Figure~\ref{fig15}. This aspect depends critically upon the specific 
emission mechanism, so one has to derive a functional relation each for the CR, ICS and SR 
mechanisms. Although this is a challenging task, it would help to identify which of the
three emission mechanisms is operating in the Crab pulsar.

In the alternate case that the second peak is not a caustic, one has to explain the 
significant rise and fall of X-ray flux as a function of phase, coupled with the monotonic
increase in $\Gamma$ as a function of phase.

In either case, Figure~\ref{fig15} contains important constraints for the process of
caustic formation in the Crab pulsar's X-ray FLC.

The third alternative, that second peak is indeed a caustic with an entirely different
PRS, will be discussed in the next section.

\subsection{Caustics with different PRS}

There is probably some observational evidence that there are at least two kinds of caustics, 
each with a different PRS. \cite{DeCesar2013} show the FLC and PRS of three RPPs that have
strong double peaks at $\gamma$-ray energies, obtained from the {\it{Fermi}} observatory;
see also \cite{Brambilla2015}. In their Figure $4.3$ the shape of the FLC and PRS ($\Gamma$)
of the Crab pulsar is roughly similar to that in Figure~\ref{fig14} above, except for 
differences in phase resolution, photon statistics, etc., and a small dip in the $\Gamma$ 
variation in the first peak. In their Figure $4.2$ the FLC of the Vela pulsar has two sharp 
peaks, but the variation of $\Gamma$ with phase appears to be approximately a phase reversed 
version of the $\Gamma$ variation in Figure~\ref{fig14}. Specifically, $\Gamma$ decreases 
monotonically with phase in the first peak, while it is almost constant in the second peak, 
which is opposite to the $\Gamma$ variation in Figure~\ref{fig14} above. In their Figure 
$4.4$ the Geminga pulsar also has two sharp peaks, and the $\Gamma$ decreases monotonically 
with phase in the first peak, just as in the Vela pulsar, while its variation is irregular 
in the second peak, but is not inconsistent with being constant. Both Vela and Geminga RPPs
have two sharp peaks each, presumably caustics, and one of them has a roughly constant $\Gamma$
with phase while the other has a monotonically decreasing $\Gamma$ with phase.

Unfortunately the PRS of these RPPs have not been obtained at soft X-ray energies. Moreover,
the FLC of Vela and Geminga RPPs have no sharp peak at soft X-ray energies, and Vela has only 
one sharp peak at hard X-ray energies \citep{Thompson2008}. Therefore this discussion has to
be done at $\gamma$-ray energies until high resolution FLC and PRS of the Vela and Geminga
RPPs is obtained at soft X-ray energies.

If the observations discussed in this section are correct, then the Vela and Geminga RPPs may have
two kinds of caustics, one with a roughly constant $\Gamma$ and the other with $\Gamma$ 
varying monotonically with phase. Then the second peak of the Crab pulsar in Figure~\ref{fig15} 
could very well be a caustic with a monotonically changing $\Gamma$ with phase. However its FLC
is not at all sharp, which is a problem from the point of the definition of a caustic.

Clearly the PRS at the second peak of the Crab pulsar in Figure~\ref{fig15} is an important
constraint for the formation of high energy caustics in RPPs.

}}

\subsection{General formation of caustics or cusps}

First, it is not clear if the sharp peak of the FLC of the Crab pulsar is on account of one or 
two poles. This can be studied by theoretically estimating the FLC and the PRS at the peak, and 
comparing it with the observations in Figure~\ref{fig15}. In the two pole scenario, one can expect 
photons from the pole farther away from the observer to traverse a region of enhanced magnetic 
field and particle density, before reaching the observer, as compared to photons from the closer 
pole, since the trajectory of the former might pass closer to the center of the RPP for 
reasonable values of $\alpha$ and $\zeta$. This might imply enhanced optical depth, either in 
terms of passage through enhanced density of plasma, or in terms of scattering off enhanced 
density of relativistic particles and photons. If this is true, then the PRS observed from two 
poles would be different from that observed from one pole only; it is likely that the two pole 
PRS would have a curvature, or even a break, in the spectrum, instead of being a simple power 
law. It is also likely that such an effect, if at all it exists, may be too weak to be observed 
by current instruments. Further, the FLC from two poles may be more luminous than that from a
single pole, but it may be broader (less sharp), since it requires alignment in phase of a
much larger volume of emission. Currently the above arguments can only be tested statistically,
since one does not have access to a FLC and PRS from a single pole that can be used as a reference.

Second, it is not clear if the sharp peak of the FLC of the Crab pulsar is formed due to the SMS 
effect. If so, then photons from the same magnetic field line, that is stretched radially, will 
arrive in phase independent of the length of the field line involved, at least to first order. 
Now, the height of the peak is proportional to the length of the field line involved 
\citep{Bai2010}, while its width may not change much since it is just one field line (or a 
narrow group of field lines). Thus in caustics formed by the 
SMS effect, the width of the peak may be relatively narrow and independent of its height. By 
the same argument, the variation of the PRS as a function of phase might be minimum. In 
contrast, a caustic formed otherwise would get contributions from a substantial volume of the 
magnetosphere, possibly with different PRS and different peak widths; in this case, the width 
of the peak may be anti correlated with its height.

Future polarization observations should show that caustics formed by the SMS effect have a 
higher degree of polarization, than caustics formed otherwise, where one can expect significant 
de-polarization due to the coming together of emission from different magnetic field lines.
{{
This has been observed in the Crab pulsar's optical FLC, where the degree of polarization at
the optical peaks is a minimum. This occurs because radiation with different polarization angles
piles up at the peaks (caustics; \cite{Smith1988, Dykes2003}; see also Figure 5 of 
\cite{Romani1995}).
}}

\subsection{Pitch angles of radiating particles}

The observed width of the sharp peak of the FLC of the Crab pulsar contains interesting 
information. First, it could be the natural width of the caustic, in which case it contains 
information about the details of the caustic formation. Alternately, the caustic itself may 
be much narrower and the observed width may be due to, say, the pitch angles of the radiating
particles in the magnetic field. In which case it contains information about the details of 
the radiative process. A proper estimate of the width is only possible if the FLC is obtained 
with sufficient resolution in rotation phase. By fitting a Gaussian to the phase range $0.395 
- 0.409$ in the top left panel of Figure~\ref{fig15}, the approximate width of the caustic 
turns out to be $0.026 \pm 0.011$ in phase. A much more accurate width estimation requires 
proper modeling of the caustic in terms of more diverse functions than a Gaussian. This can 
also be done by complete FLC modeling at the peak.

\begin{acknowledgements}
I thank the referee for bringing to my attention the IACHEC reports, and the referee and 
Sergio Campana for useful discussion and suggestions.
\end{acknowledgements}

\begin{appendix}
\section{High Background counts}

First one extracts from the event file a LTLC in the energy range $12 - 15$ keV (energy 
channels $1200 - 1500$) with a bin width of $8$ s, using the {\it{extractor}} tool. Next 
one identifies those time bins in the LTLC that have count rates greater than $1.0$. Let 
these be TTIME. These times represent the center of the time bin if the keyword TIMEPIXR 
$= 0.5$; the bin width is given in the keyword TIMEDEL. The start time of the LTLC is given 
in the keyword TIMEZERO, which represents the center of the first time bin. Finally the 
keyword of the same name, TIMEZERO, but now in the event file, contains a clock correction 
that is usually $-1$ s; here it will be labeled CLKCOR. Therefore the relevant time 
range to be excluded in the event file (in units of TT) for each TTIME is given by 
TIMEZERO + TTIME - CLKCOR $\mp$ TIMEDEL$/2$, representing the lower and upper limits of
the duration to be excluded, respectively.

The above time ranges have to be excluded form the GTI in the event file, in the FITS file 
extension $10$. There are several ways of doing this. But an elegant way is to use the 
algorithm given in the FORTRAN function {\it{gtimerge}} in the library 
{\it{gtilib.f}}\footnote{/apln/heasoft-6.27.2/ftools/ftoolslib/gen/gtilib.f}.

\section{Space weather parameter Kp}

The official Kp web page is at the German Research Center for Geosciences (GFZ);
the link is given in footnote $1$ of this paper. The WDC directory contains the file 
{\it{wdc\_fmt.txt}} which contains the format of the data in the four ASCII files of use 
here, viz., {\it{kp2017.wdc}}, {\it{kp2018.wdc}}, {\it{kp2019.wdc}} and {\it{kp2020.wdc}},
one for each year of operation of {\it{NICER}}, from which the Kp data has to be retrieved 
for each ObsID.

One begins by extracting the start and end times of the observation from the event file 
for each ObsID, from the keywords TSTART and TSTOP. One also reads the integer and 
fractional parts of the reference MJD for {\it{NICER}} (MJDREFI and MJDREFF), and the clock 
correction TIMEZERO (this is labeled CLKCOR in Appendix A). Then the start and end times 
are converted into MJD using the formula TT(1,2) = MJDREFI + MJDREFF + ((TSTART,TSTOP) + 
TIMEZERO) / 86400, where TT1 and TT2 correspond to TSTART and TSTOP, respectively. These 
are converted into Gregorian dates using the SOFA\footnote{/apln/heasoft-6.27.2/heacore/ast/sofa/} 
software {\it{iauTttai}}, {\it{iauTaiutc}}, {\it{iauJd2cal}}, which convert the start 
and end times of observations first from TT in MJD units to TAI in MJD, then from TAI 
to UTC, then from UTC in JD units to Gregorian calendar date, respectively; note the 
conversion from MJD to JD before the last step. The conversion to date has been 
verified using the web tool {\it{xTime}}\footnote{https://heasarc.gsfc.nasa.gov/cgi-bin/Tools/xTime/xTime.pl}, 
a date and time conversion utility.

Based on the year of observation, the corresponding WDC file is read, and the Kp is plotted
as a function of time during the observation; Kp is given over $3$ hour intervals. If a 
Kp value is greater than $5$, then that entire $3$ hours of observation has to be excluded.

\section{Linux shell scripts to obtain and analyze PRS}

Obtaining and analyzing manually the spectral data of Figures \ref{fig14} and \ref{fig15} 
is difficult. So Linux shell scripts were developed to do the tasks automatically.

First, the FLC of the top panels of these figures is obtained, using a C/C++ program.
The output of this is an ASCII file containing two columns, viz, the central phase and 
the photon counts of each bin in rotation phase; in Figures \ref{fig14} that would be
$128$ bins.

Next a Linux shell script (in {\it{tcsh}}) reads the column of central phases, computes
the phase boundaries of the corresponding bins, and invokes the {\it{extractor}} tool
to obtain the spectrum of the data within each phase bin, using $27$ ObsIDs. The output 
of this script are FITS files of type {\it{pha}}, one for each phase bin. These are the 
so called PRS that can be analyzed using XSPEC. Next a Linux shell script (in {\it{tcsh}}) 
writes the required XSPEC commands into an ASCII file with extension {\it{.xcm}}, which 
is invoked as ``xspec - *.xcm''.

The first XSPEC command defines data groups; in Figure~\ref{fig14} it would be $78$ data 
groups, one for each {\it{pha}} file being analyzed. The command would be ``data 1:1 
y\_38.pha 2:2 y\_39.pha ... 78:78 y\_115.pha'', the file names indicating the 
corresponding phase bin; in Figure~\ref{fig14}, phase bins $38$ to $115$ have been
analyzed. The next two commands set the response and ancillary response files for each 
data group: ``response 1 nixti20170601\_combined\_v002.rmf 2 nixti20170601\_combined\_v002.rmf 
... 78 nixti20170601\_combined\_v002.rmf'' and ``arf 1 nixtionaxis20170601\_combined\_v004.arf 
2 nixtionaxis20170601\_combined\_v004.arf ... 78 nixtionaxis20170601\_combined\_v004.arf''.

Next the background spectrum file is defined for each group, as well as the channels
to ignore for each group. Next the model to fit is defined ``model phabs(powerlaw)'', 
followed by the command ``/*''. Then the initial values of the parameters to fit are
defined using the ``newpar'' command. Since there are $78$ spectra each with $3$
parameters, there are totally $3 \times 78 = 234$ parameters to fit. However the
first parameter of each spectrum $N_H$ is fixed at the value $0.36$. So the next
$78$ commands are ``newpar 1 0.36; newpar 4 = 1; newpar 7 = 1; ...  newpar 232 = 1'';
The command ``freeze 1'' ensures that these $78$ parameters are not fit but held 
fixed. Each command must be on a separate line. Similarly the second and third 
parameters of the model ($\Gamma$ and $A$) are given initial values, but they are 
not frozen. Then the following three commands result in the fitting process: 
``renorm; query yes; fit'', each on a separate line.

The ``ignore'' and ``model'' commands require the use of the star symbol ``*'',
which is easy to issue manually in XSPEC but tricky to do so in a shell script. 

Finally the command to estimate the $90$\% confidence errors on the parameters
is:``error 2 3 5 6 8 9 ... 230 231 233 234'' on one line. Note that the parameters 
1, 4, 7, ..., 229, 232 do not appear in the command, since they refer to the 
parameter $N_H$ which is not fit. To speed up the execution of this command, the
command ``parallel error 4'' uses $4$ cores of the computer's CPU for parallel 
processing.

The error command requires a lot of time to execute. The whole {\it{xcm}} script 
for $128$ phase bins requires $\approx 3.0$ days of continuous running on an 
average laptop -- Acer Aspire E5-575, Intel core i$3$-$7100$U CPU @ $2.40$GHz, 
$2$ cores, $4$ threads, $12$ GB DDR4 Synchronous RAM @ $2133$ MHz, $64$ bit 
Fedora core $32$ operating system. The script would require $\approx 3.5$ 
years of continuous running for $1024$ phase bins.

\end{appendix}

\begin{thebibliography}{}
\bibitem[Abdo et al. (2010)]{Abdo2010} Abdo, A.A., Ackermann, M., Ajello, M. et al. 2010, \apj, 708, 1254
\bibitem[Arzoumanian et al. (2014)]{Arzoumanian2014} Arzoumanian, Z., Gendreau, K.C., Baker, C.L., et al. 2014, Proc. SPIE, 9144, 914420
\bibitem[Atwood et al. (2007)]{Atwood2007}Atwood, W. B., Bagagli , R., Baldini , L., et al. 2007, Astro. Phys. 28, 422
\bibitem[Bai \& Spitkovsky (2010)]{Bai2010} Bai, X.N., Spitkovsky, A., 2010, \apj, 715, 1282
\bibitem[Bogdanov et al. (2019)]{Bogdanov2019} Bogdanov, S., Ho, W.C.G., Enoto, T. et al. 2019, \apj, 877, 69
\bibitem[Brambilla et al. (2014)]{Brambilla2015} Brambilla, G., Kalapotharakos, C., Harding, A.K., et al. 2015, \apj, 804, 34
\bibitem[Bult et al. (2018)]{Bult2018} Bult, P., Altamirano, D., Arzoumanian, Z. et al. 2018, \apjl, 860, L9
\bibitem[Cheng et al. (1986)]{Cheng1986} Cheng, K.S., Ho, C. \& Ruderman, M. 1986, \apj, 300, 500
\bibitem[Cheng et al. (2000)]{Cheng2000} Cheng, K.S., Ruderman, M. \& Zhang, L. 2000, \apj, 537, 964
\bibitem[DeCesar (2013)]{DeCesar2013} DeCesar, M. 2013, PhD thesis, Univ. Maryland
\bibitem[Dykes \& Rudak (2013)]{Dykes2003} Dykes, J. \& Rudak, B. 2003, \apj, 598, 1201
\bibitem[Ge et al. (2012)]{Ge2012} Ge, M.Y., Lu, F.J., Qu, J.L. et al., 2012, \apjs, 199, 32
\bibitem[Gendreau et al. (2016)]{Gendreau2016}Gendreau, K., Arzoumanian, Z., Adkins, P.W., et al. 2016, Proc. SPIE, 9905, 99051H 
\bibitem[Guillot et al. (2019)]{Guillot2019} Guillot, S.,  Kerr, M., Ray, P.S. et al. 2019, \apjl, 887, L27
\bibitem[Harding (2016)]{Harding2016} Harding, A. 2016, J. Plasma Phys. 82, 635820306
\bibitem[Hare et al. (2020)]{Hare2020} Hare, J., Tomsick, J.A.,  Buisson, D.J.K. et al. 2020, \apj, 890, 57
\bibitem[Kalapotharakos et al. (2014)]{Kalapotharakos2014} Kalapotharakos, C., Harding, A.K., \& Kazanas, D. 2014, \apj, 793, 97
\bibitem[Keek et al. (2018)]{Keek2018} Keek, L., Arzoumanian, Z., \& Bult, P. 2018, \apjl, 855, L4
\bibitem[Kirsch et al. (2005)]{Kirsch2005} Kirsch, M.G.F., Briel, U.G., Burrows, D. et al. 2005, Proc. of SPIE Vol. 5898,
	"UV, X-Ray, and Gamma-Ray Space Instrumentation for Astronomy", Eds O. H. W. Siegmund, 589803-1
\bibitem[Kirsch et al. (2006)]{Kirsch2006} Kirsch, M.G.F., Schönherr, G., Kendziorra, E. et al. 2006, \aap, 453, 173
\bibitem[Kouzu et al. (2013)]{Kouzu2013} Kouzu, T., Tashiro, M.S., Terada, Y. et al. 2013, \pasj, 65, 74
\bibitem[Ludlam et al. (2018)]{Ludlam2018} Ludlam, R.M., Miller, J.M., Arzoumanian, Z. et al. 2018, \apjl, 858, L5
\bibitem[Ludlam et al. (2019)]{Ludlam2019} Ludlam, R.M., Shishkovsky, L., Bult, P.M. et al. 2019, \apj, 883, 39
\bibitem[Madsen et al. (2015)]{Madsen2015} Madsen, K. K., Reynolds, S., Harrison, F., et al. 2015, \apj, 801, 66
\bibitem[Malacaria et al. (2019)]{Malacaria2019} Malacaria, C., Bogdanov, S., Ho, W.C.G., et al. 2019, \apj, 880, 74
\bibitem[Massaro et al. (2000)]{Massaro2000} Massaro, E., Cusumano, G., Litterio, M., \& Mineo, T. 2000, \aap, 361, 695
\bibitem[Massaro et al. (2006)]{Massaro2006} Massaro, E., Campana, R., Cusumano, G., \& Mineo, T. 2006, \aap, 459, 859
\bibitem[Miller et al. (2019)]{Miller2019} Miller, J.M., Kammoun, E., Ludlam, R.M.,  et al. 2019, \apj, 884, 106
\bibitem[Pierbattista et al. (2016)]{Pierbattista2016} Pierbattista, M., Harding, A.K., Gonthier, P.L. et al. 2016, \aap, 588, A137
\bibitem[Pravdo et al. (1997)]{Pravdo1997} Pravdo, S.H.  Angelini, L., \& Harding, A.K. 1997, \apj, 491, 808
\bibitem[Prigozhin et al. (2016)]{Prigozhin2016} Prigozhin, G., Gendreau, K., Doty, J.P., et al. 2016, Proc. SPIE, 9905, 99051I
\bibitem[Ray et al. (2019)]{Ray2019} Ray, P.S., Guillot, S., Ransom, S.M., et al. 2019, \apj, 878, L22
\bibitem[Riley et al. (2019)]{Riley2019} Riley, T.E., Watts, A.L., Bogdanov, S. et al. 2019, \apjl, 887, L21
\bibitem[Romani \& Yadigaroglu (1995)]{Romani1995} Romani, R.W. \& Yadigaroglu, I.A. 1995, \apj, 438, 314
\bibitem[Romani \& Watters (2010)]{Romani2010} Romani, R.W. \& Watters, K.P. 2010, \apj, 714, 810
\bibitem[Smith et al. (1988)]{Smith1988} Smith, F. G., Jones, D.H.P., Dick, J.S.B. et al. 1988, \mnras, 233, 305
\bibitem[Stevens et al. (2019)]{Stevens2018} Stevens, A.L., Uttley, P., Altamirano, D. et al. 2018, \apjl, 865, L15
\bibitem[Thompson (2008)]{Thompson2008} Thompson, D. J. 2008, Reports on Progress in Physics, 71, 116901
\bibitem[Trakhtenbrot et al. (2019)]{Trakhtenbrot2019} Trakhtenbrot, B., Arcavi, I., Ricci, C. et al. 2019, Nature Astronomy, 3, 187
\bibitem[Tuo et al. (2019)]{Tuo2019} Tuo, Y.L., Ge, M.Y., Song, L.M. et al. 2019, RAA, 19, 87
\bibitem[van den Eijnden et al. (2020)]{vandenEijnden2020} van den Eijnden, J., Degenaar, N., Ludlam, R.M. et al. 2020, \mnras, 493, 1318
\bibitem[Vivekanand (2020)]{Vivekanand2020} Vivekanand, M. 2020, \aap, 633, A57
\bibitem[Weisskopf et al. (2011)]{Weisskopf2011} Weisskopf, M.C., Tennant, A.F., Yakovlev, D.G. et al. 2011, \apj, 743, 139
\bibitem[Wilson-Hodge et al. (2011)]{Wilson-Hodge2011} Wilson-Hodge, C.A., Cherry, M.L., Case, G.L. et al. 2011, \apj, 727, L40
\bibitem[Wilson-Hodge et al. (2018)]{Wilson-Hodge2018} Wilson-Hodge, C.A., Malacaria, C., Jenke, P.A. et al. 2018, \apj, 863, 9
\end{thebibliography}
\end{document}